\documentclass[final,3p,times]{elsarticle}
\usepackage{multirow,setspace,times,amssymb,amsmath,graphicx,color,rotating,subfigure,url}
\usepackage{lineno}
\usepackage{dcolumn,multirow}
\usepackage{natbib}

\journal{Physica A} 

\begin{document}

\begin{frontmatter}

\title{Empirical properties of inter-cancellation durations in the Chinese stock market}
\author[SB,RCE]{Gao-Feng Gu}
\author[CME,CCSCA]{Xiong Xiong}
\author[CME,CCSCA]{Wei Zhang}
\author[CME,CCSCA]{Yong-Jie Zhang}
\author[SB,RCE,SS]{Wei-Xing Zhou\corref{cor}}
\cortext[cor]{Corresponding author. Address: 130 Meilong Road, P.O. Box 114, School of Business, East China University of Science and Technology, Shanghai 200237, China, Phone: +86 21 64253634, Fax: +86 21 64253152.}
\ead{wxzhou@ecust.edu.cn} %

\address[SB]{School of Business, East China University of Science and Technology, Shanghai 200237, China}
\address[RCE]{Research Center for Econophysics, East China University of Science and Technology, Shanghai 200237, China}
\address[CME]{College of Management and Economics, Tianjin University, Tianjin 300072, China}
\address[CCSCA]{China Center for Social Computing and Analytics, Tianjin University, Tianjin 300072, China}
\address[SS]{School of Science, East China University of Science and Technology, Shanghai 200237, China}

\begin{abstract}
Order cancellation process plays a crucial role in the dynamics of price formation in order-driven stock markets and is important in the construction and validation of computational finance models. Based on the order flow data of 18 liquid stocks traded on the Shenzhen Stock Exchange in 2003, we investigate the empirical statistical properties of inter-cancellation durations in units of events defined as the waiting times between two consecutive cancellations. The inter-cancellation durations for both buy and sell orders of all the stocks favor a $q$-exponential distribution when the maximum likelihood estimation method is adopted; In contrast, both cancelled buy orders of 6 stocks and cancelled sell orders of 3 stocks prefer Weibull distribution when the nonlinear least-square estimation is used. Applying detrended fluctuation analysis (DFA), centered detrending moving average (CDMA) and multifractal detrended fluctuation analysis (MF-DFA) methods, we unveil that the inter-cancellation duration time series process long memory and multifractal nature for both buy and sell cancellations of all the stocks. Our findings show that order cancellation processes exhibit long-range correlated bursty behaviors and are thus not Poissonian.
\end{abstract}


\end{frontmatter}

\section{Introduction}
\label{se:introduction}

In an order-driven market, order submission and cancellation play the most important role in the process of price formation. For order submission process, lots of studies have been conducted to investigate the statistical properties of the ingredients of an order including order price \citep{Zovko-Farmer-2002-QF,Bouchaud-Mezard-Potters-2002-QF,Potters-Bouchaud-2003-PA,Ranaldo-2004-JFinM,Maskawa-2007-PA,Lillo-2007-EPJB,Mike-Farmer-2008-JEDC,Gu-Chen-Zhou-2008b-PA}, order size or volume \cite{Gopikrishnan-Plerou-Gabaix-Stanley-2000-PRE,Plerou-Gopikrishnan-Gabaix-Amaral-Stanley-2001-QF,Farmer-Lillo-2004-QF,Queiros-2005-EPL,deSouza-Moyano-Queiros-2006-EPJB,Plerou-Stanley-2007-PRE,Mu-Chen-Kertesz-Zhou-2009-EPJB,Gu-Ren-Ni-Chen-Zhou-2010-PA,Zhou-2012-NJP,Zhou-2012-QF}, order direction \cite{Mike-Farmer-2008-JEDC,Lillo-Farmer-2004-SNDE,Gu-Zhou-2009-EPL,Gu-Zhou-2009-EPJB}, and so on. Special attention has been paid to the probability distribution and memory effect of these ingredients and many stylized facts have been documented.

Order cancellation is a process of removing orders from the limit-order book which is a queue of limit orders waiting to be executed and constructed according to the price-time priority. If all orders placed at the best ask or best bid are cancelled, the mid-price defined as the mean value of the best ask and best bid will change. If cancellation occurs inside the limit order book, it causes changes of the structure of limit order book and has potential impact on price fluctuation.

The motivation of order cancellation is related to the non-execution (NE) risk or free option (FO) risk \cite{Griffiths-Smith-Turnbull-White-2000-JFE,Fong-Liu-2010-JBF}, and the former is the major reason for cancelling limit orders \cite{Fong-Liu-2010-JBF}. NE risk arises when the current security price moves away from the submitting price. The orders submitted in the front of limit-order book cannot be transacted immediately, which makes the traders suffer opportunity cost. Traders may cancel the stale orders and resubmit more aggressive ones to increase the transaction probability. So in order to reduce NE risk, buy traders potentially drive the security price up, and sell traders drive the price down. FO risk arises when important news arrives. The intrinsic value of asset will be underestimated (for good news) or overestimated (for bad news) according to the current price. In order to prevent to be traded at the unfavorable price, traders will cancel their limit orders and resubmit unaggressive ones. So conversely, in order to reduce FO risk, buy traders potentially drive the price down, and sell traders drives the price up.

Since there are rare cancellation data recorded in the past, only a few literatures investigated the empirical regularities of order cancellation. With the development of information technology and computer science, it is possible to record the order flow data which enables us to analyze the statistical properties of order cancellation and construct cancellation models. Ni et al. investigated the empirical regularities of inter-cancellation duration of 22 stocks in the Chinese stock market, and made a conclusion that order cancellation is a non-Poisson process \cite{Ni-Jiang-Gu-Ren-Chen-Zhou-2010-PA}. Liu showed a simple model of order revision and cancellation, and found that the frequency of order cancellation is positively related to order submission risk and stock capitalization, but negatively related to bid-ask spread \cite{Liu-2009-JFM}. In an order-driven model, Daniels et al. assumed that order cancellation follows a Poisson process, which makes the model having powerful predictions of stylized facts, such as price diffusion, price impact, and so on \cite{Daniels-Farmer-Gillemot-Iori-Smith-2003-PRL}. In the empirical model proposed by Mike and Farmer, the order cancellation process is determined by three independent factors, the position in the order book relative to the opposite best price, the imbalance of buy and sell orders in the limit-order book, and the total number of orders stored in the limit-order book. This cancellation model gives an excellent prediction for the life time of cancelled orders \cite{Mike-Farmer-2008-JEDC}.

In financial markets, a widely studied subject is the recurrence interval defined as the waiting time between two consecutive events. Many scholars have analyzed the probability distribution of recurrence intervals of different variables such as returns, volatilities and trading volumes. However, the results are controversial. Power-law distribution \cite{Kaizoji-Kaizoji-2004a-PA,Yamasaki-Muchnik-Havlin-Bunde-Stanley-2005-PNAS,Lee-Lee-Rikvold-2006-JKPS,Ren-Zhou-2010-PRE} and stretched exponential distribution \cite{Wang-Yamasaki-Havlin-Stanley-2006-PRE,Wang-Weber-Yamasaki-Havlin-Stanley-2007-EPJB,VodenskaChitkushev-Wang-Weber-Yamasaki-Havlin-Stanley-2008-EPJB,Jung-Wang-Havlin-Kaizoji-Moon-Stanley-2008-EPJB,Wang-Yamasaki-Havlin-Stanley-2008-PRE,Qiu-Guo-Chen-2008-PA,Ren-Zhou-2008-EPL,Ren-Gu-Zhou-2009-PA,Ren-Guo-Zhou-2009-PA,Wang-Wang-2012-CIE,Meng-Ren-Gu-Xiong-Zhang-Zhou-Zhang-2012-EPL,Xie-Jiang-Zhou-2014-EM}
are mainly selected to fit the probability density function (PDF) in different financial markets. Moreover, other distributions are also proposed for complement \cite{Zhang-Wang-Shao-2010-ACS,Jeon-Moon-Oh-Yang-Jung-2010-JKPS}. It is interesting that the recurrence interval time series usually processes long memory \cite{Yamasaki-Muchnik-Havlin-Bunde-Stanley-2005-PNAS,Ren-Zhou-2010-PRE,Wang-Yamasaki-Havlin-Stanley-2006-PRE,Wang-Weber-Yamasaki-Havlin-Stanley-2007-EPJB,Jung-Wang-Havlin-Kaizoji-Moon-Stanley-2008-EPJB,Ren-Gu-Zhou-2009-PA,Ren-Guo-Zhou-2009-PA,Meng-Ren-Gu-Xiong-Zhang-Zhou-Zhang-2012-EPL,Xie-Jiang-Zhou-2014-EM,Ren-Zhou-2010-NJP,Yamasaki-Muchnik-Havlin-Bunde-Stanley-2006-inPFE}.
Recurrence interval analysis has also been applied to other fields such as the energy dissipation rate in three-dimensional fully developed turbulence \cite{Liu-Jiang-Ren-Zhou-2009-PRE}.

Similar to recurrence interval, intertrade duration which is defined as the interval between two consecutive transactions is another important topic. Continuous-time random walk (CTRW) proposed by Montroll and Weiss \cite{Montroll-Weiss-1965-JMP} has been widely utilized to deal with the intertrade duration in the financial time series \cite{Scalas-Gorenflo-Mainardi-2000-PA,Mainardi-Raberto-Gorenflo-Scalas-2000-PA,Masoliver-Montero-Weiss-2003-PRE,Kim-Yoon-2003-Fractals,Scalas-2006-PA,Masoliver-Montero-Perello-Weiss-2006-JEBO}. Empirical results indicate that intertrade durations might follow a power-law distribution \cite{Sabatelli-Keating-Dudley-Richmond-2002-EPJB,Masoliver-Montero-Weiss-2003-PRE,Masoliver-Montero-Perello-Weiss-2006-JEBO,Yoon-Choi-Lee-Yum-Kim-2006-PA}, a stretched exponential or Weibull distribution \cite{Raberto-Scalas-Mainardi-2002-PA,Bartiromo-2004-PRE,Ivanov-Yuen-Podobnik-Lee-2004-PRE,Eisler-Kertesz-2006-EPJB,Sazuka-2007-PA,Jiang-Chen-Zhou-2008-PA}, or a $q$-exponential distribution \cite{Jiang-Chen-Zhou-2008-PA,Poloti-Scalas-2008-PA}. On the other hand, some studies showed that the intertrade durations are neither exponentially distributed \cite{Sabatelli-Keating-Dudley-Richmond-2002-EPJB,Sazuka-2007-PA,Scalas-Gorenflo-Luckock-Mainardi-Mantelli-Raberto-2004-QF,Scalas-Gorenflo-Luckock-Mainardi-Mantelli-Raberto-2005-FL}
nor power-law distributed \cite{Poloti-Scalas-2008-PA}.

The goodness of fit for Weibull distribution and $q$-exponential distribution has been estimated in the distribution of intertrade durations. Jiang et al. studied the limit order data of 18 liquid stocks listed in the Chinese stock market, and showed that Weibull distribution gives better fitting than $q$-exponential distribution with the maximum likelihood estimation method, while $q$-exponential distribution outperforms Weibull distribution with the nonlinear least-squares estimation method \cite{Jiang-Chen-Zhou-2008-PA}. Poloti and Scalas analyzed the tick-by-tick data set of DJIA stocks traded at NYSE in year 1999, and found that $q$-exponential distribution compares well to the Weibull distribution \cite{Poloti-Scalas-2008-PA}.

In this paper, we will study the statistical properties of inter-cancellation durations in event time for both cancelled buy and sell orders of 18 stocks listed on the Shenzhen Stock Exchange. The rest of paper is organized as follows. We study the probability distributions of the inter-cancellation durations based on the maximum likelihood estimation and nonlinear least-square estimation methods. We further discuss the memory effect and multifractal nature.

\section{Dateset}
\label{se:dateset}

Our analysis is based on the order flow data of 18 liquid stocks traded on the Shenzhen Stock Exchange in 2003. There were three periods in a trading day in 2003: opening call auction, cool period and continuous double auction. Opening call auction is held from 9:15 a.m. to 9:25 a.m., referring to the process of one-time centralized matching of buy and sell orders accepted during a specified period to generate the opening price at 9:25 in a trading day. Following the opening call auction, cool period is held from 9:25 a.m. to 9:30 a.m. when the Exchange is opened to orders routing from investors, but the orders or cancellations are not allowed to be processed. The main trading period is the continuous auction (9:30 a.m. - 11:30 a.m. and 13:00 p.m. - 15:00 p.m.), which refers to the process of continuous matching of both buy and sell orders on a one-by-one basis.

\begin{table}[htp]
  \centering
  \caption{The statistics of inter-cancellation durations for both cancelled buy and sell orders of 18 stocks. $N_C$ is the number of cancellation. $r$ is the ratio of $N_C$ to $N_A$. $\gamma$ is the slope of the fitted line presented in the figure below. $\langle{d}\rangle$ is the average inter-cancellation duration in units of events.}
  \label{Tb:Dataset-Number}
  \centering
  \begin{tabular}{ccrccrrrccr}
  \hline \hline
  Stock && \multicolumn{4}{c}{Cancelled buy orders} && \multicolumn{4}{c}{Cancelled sell orders} \\
  \cline{3-6} \cline{8-11}
  && $N_C$~~~~ & $r$ & $\gamma$ & $\langle{d}\rangle$~~ && $N_C$~~~ & $r$ & $\gamma$ & $\langle{d}\rangle$~~ \\
  \hline
    000001 && 320 872 & 0.157 & 0.174 & 12.12 && 277 878 & 0.148 & 0.148 & 13.73 \\
    000009 && 185 018 & 0.176 & 0.180 & 11.81 && 188 725 & 0.164 & 0.183 & 11.38 \\
    000012 && 115 335 & 0.194 & 0.187 &  9.84 && 107 443 & 0.189 & 0.200 & 10.62 \\
    000016 &&  60 724 & 0.170 & 0.179 & 11.89 &&  59 723 & 0.154 & 0.152 & 12.09 \\
    000021 && 158 470 & 0.188 & 0.182 & 10.70 && 154 157 & 0.174 & 0.173 & 10.98 \\
    000024 &&  43 013 & 0.177 & 0.179 & 12.03 &&  46 130 & 0.156 & 0.160 & 11.23 \\
    000066 && 110 979 & 0.191 & 0.188 & 10.60 && 107 641 & 0.173 & 0.182 & 10.85 \\
    000406 &&  94 998 & 0.164 & 0.160 & 11.92 &&  92 841 & 0.157 & 0.173 & 12.21 \\
    000429 &&  37 332 & 0.161 & 0.158 & 12.63 &&  36 585 & 0.140 & 0.138 & 12.96 \\
    000488 &&  33 183 & 0.148 & 0.167 & 13.18 &&  34 478 & 0.145 & 0.150 & 12.75 \\
    000539 &&  27 322 & 0.127 & 0.124 & 14.76 &&  27 487 & 0.127 & 0.128 & 14.88 \\
    000541 &&  19 938 & 0.149 & 0.178 & 13.29 &&  20 112 & 0.129 & 0.135 & 13.24 \\
    000550 && 123 948 & 0.188 & 0.193 & 10.99 && 131 270 & 0.181 & 0.182 & 10.31 \\
    000581 &&  27 570 & 0.152 & 0.169 & 13.99 &&  29 907 & 0.131 & 0.124 & 13.05 \\
    000625 && 124 736 & 0.178 & 0.187 & 11.38 && 133 627 & 0.179 & 0.187 & 10.65 \\
    000709 &&  66 291 & 0.152 & 0.145 & 13.16 &&  64 697 & 0.136 & 0.145 & 13.55 \\
    000720 &&  16 767 & 0.097 & 0.116 & 17.44 &&  14 441 & 0.087 & 0.089 & 20.29 \\
    000778 &&  44 113 & 0.151 & 0.150 & 14.17 &&  48 041 & 0.131 & 0.138 & 13.13 \\
  \hline \hline
 \end{tabular}
\end{table}

Our database records ultra-high-frequency order flow data whose time stamps are accurate to 0.01s. It contains the details of order placement and order cancellation. For example, the stock Ping An Bank Co., Ltd. (000001) contains 3,925,832 records in the whole year of 2003. So we can rebuild the limit-order book based on the prefect database according the trading rules. On the other hand, it is difficult to obtain this type of prefect database and we only have the data of 23 stocks in the whole year of 2003 among which 5 stocks have wrong records of order cancellation, and we select the rest 18 stocks to study the statistical properties of order cancellation. For the 18 stocks analyzed, they cover 9 CSRC (China Securities Regulatory Commission) Industries, such as finance and insurance, real estate, transportation, machinery, to list a few. On the other hand, in the year of 2003, the Chinese stock index first went up then fell down. Bull market and bear market were both existed in 2003. So the database we studied generally presents the situation of Chinese market.

The paper not only focuses on the cancellation data in the continuous auction, but also includes the cancellation data in the opening call auction and cool period. We count the cancellation numbers $N_C$ for both cancelled buy and sell orders of each stock, and then calculate the ratio $r$ of $N_C$ to the number of all the orders $N_A$ (including both submitted orders and cancelled orders). The results are listed in Table \ref{Tb:Dataset-Number}. We find that the ratio $r$ fluctuates within a wide range, being $[0.097,0.195]$ for cancelled buy orders and being $[0.087,0.189]$ for cancelled sell orders. An interesting feature shows that the ratio of cancelled buy orders approximates to cancelled sell orders for each stock, which implies that a large proportion of buy orders cancelled corresponds to a large number of cancellations for sell orders, and vice versa.

\begin{figure}[htb]
\centering
\includegraphics[width=7cm]{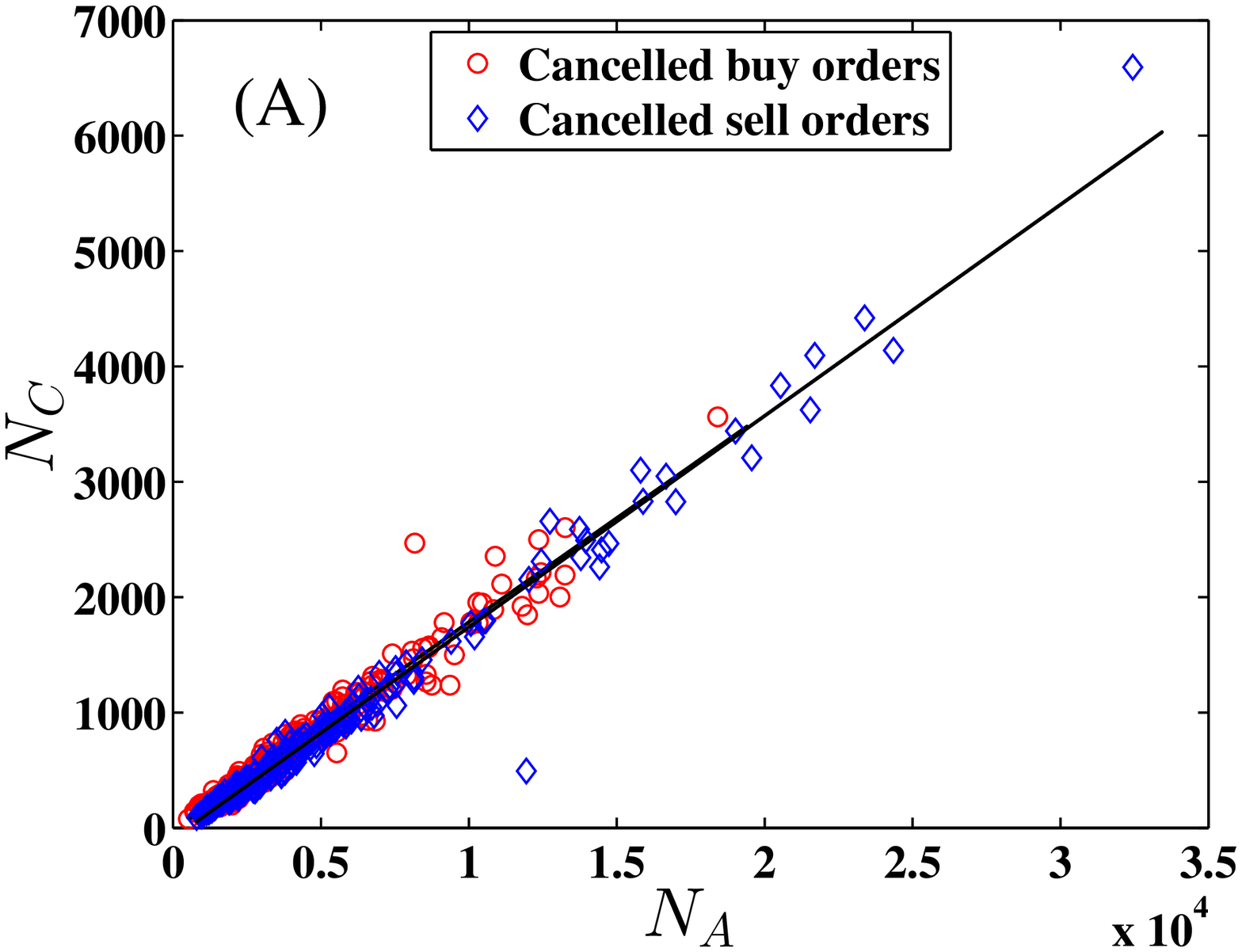}
\includegraphics[width=7cm]{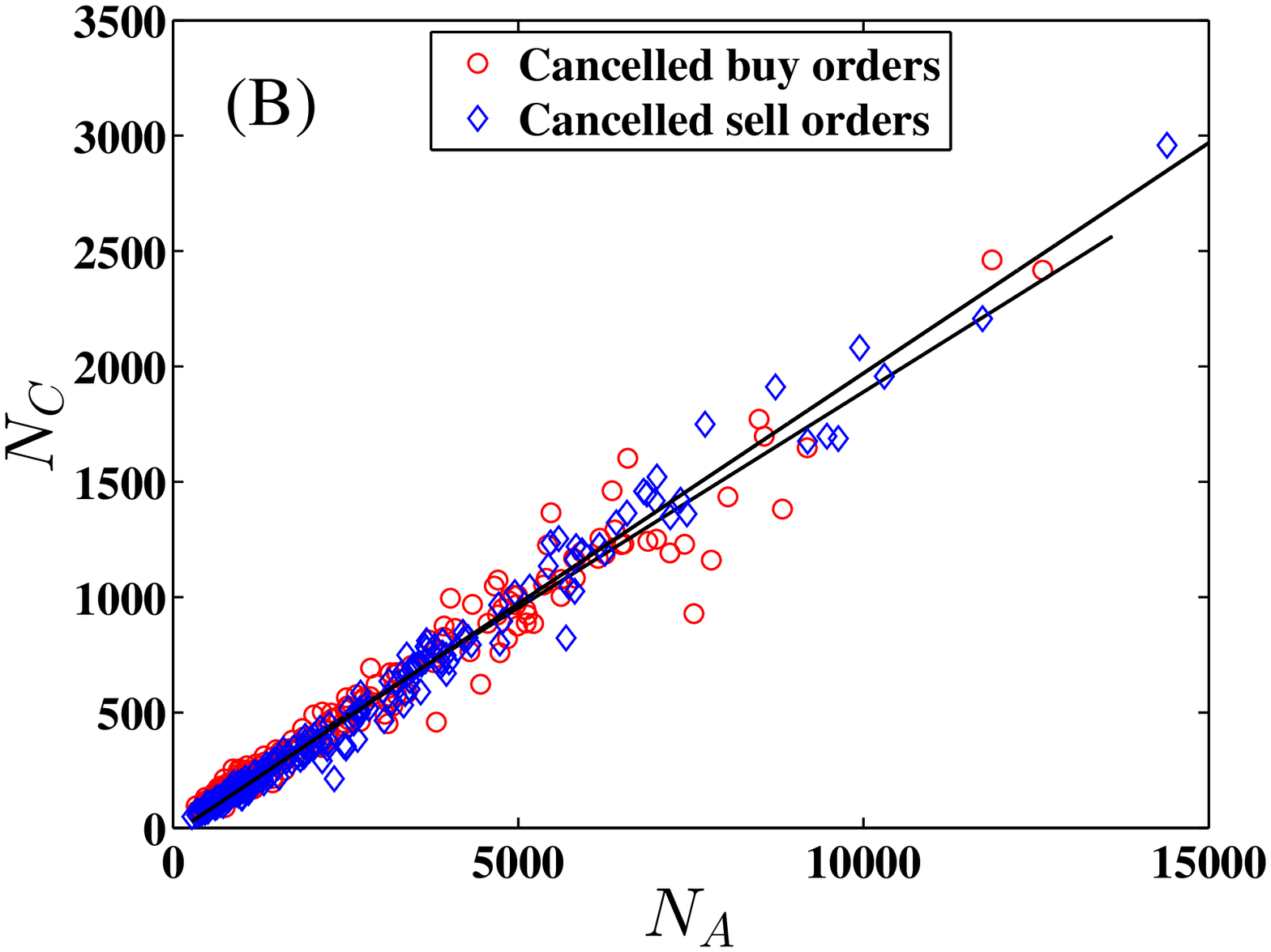}
\caption{\label{Fig:Dataset-Number} Plots of the number of cancellations $N_C$ with respect to the number of all the orders $N_A$ of each trading day for two stocks 000009 (a) and 000012 (b). The solid lines are calculated with the least-squares fitting method.}
\end{figure}

We further investigate the ratios $r$ in each trading day for all the stocks. Figure \ref{Fig:Dataset-Number} presents the linear relations between $N_C$ and $N_A$ for both cancelled buy and sell orders of two stocks. The slopes $\gamma$ of the fitted lines are calculated using the least-squares fitting method, and the values of 18 stocks are listed in Table \ref{Tb:Dataset-Number}. It is evident that the value of $\gamma$ is close to the value of $r$ for each stock, which implies that the ratios of cancellation in each trading day are almost similar for both cancelled buy and sell orders of each stock.

In the paper, the inter-cancellation duration is defined as the interval between two consecutive cancellations in units of events, which reads
\begin{equation}
  d(i)=t(i)-t(i-1)~,
  \label{Eq:df_cd}
\end{equation}
where $t(i)$ is the event time when the $i$-th cancellation takes place. It is clear that inter-cancellation duration $d(i)$ is the number of orders (including both buy orders and sell orders) submitted between the $(i-1)$-th cancellation and the $i$-th cancellation. We calculate the average values $\langle{d}\rangle$ of inter-cancellation durations for both cancelled buy and sell orders of each stock, and depict the results in Table \ref{Tb:Dataset-Number}. According to the definition of inter-cancellation duration, we easily obtain the relation $r>1/\langle{d}\rangle$ which is confirmed by the data listed in the table. The reason is that the ratio $r$ is defined based on a certain kind of orders (buy orders or sell orders) while the inter-cancellation duration $d$ is considered as the number of both buy and sell orders submitted between two successive cancellations.

\section{Probability distribution}
\label{se:PDF}

\begin{figure}[htb]
  \centering
  \includegraphics[width=7cm]{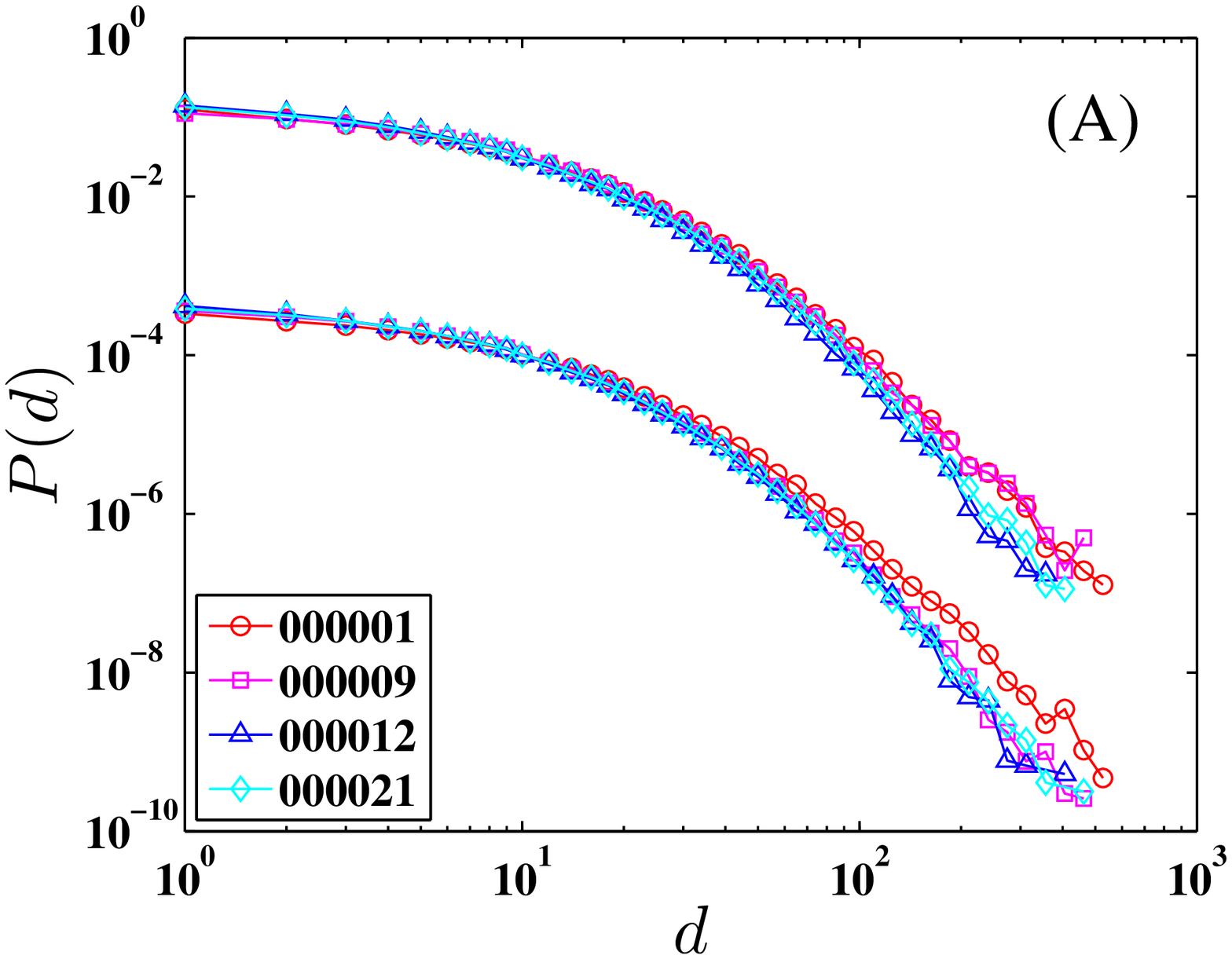}
  \includegraphics[width=7cm]{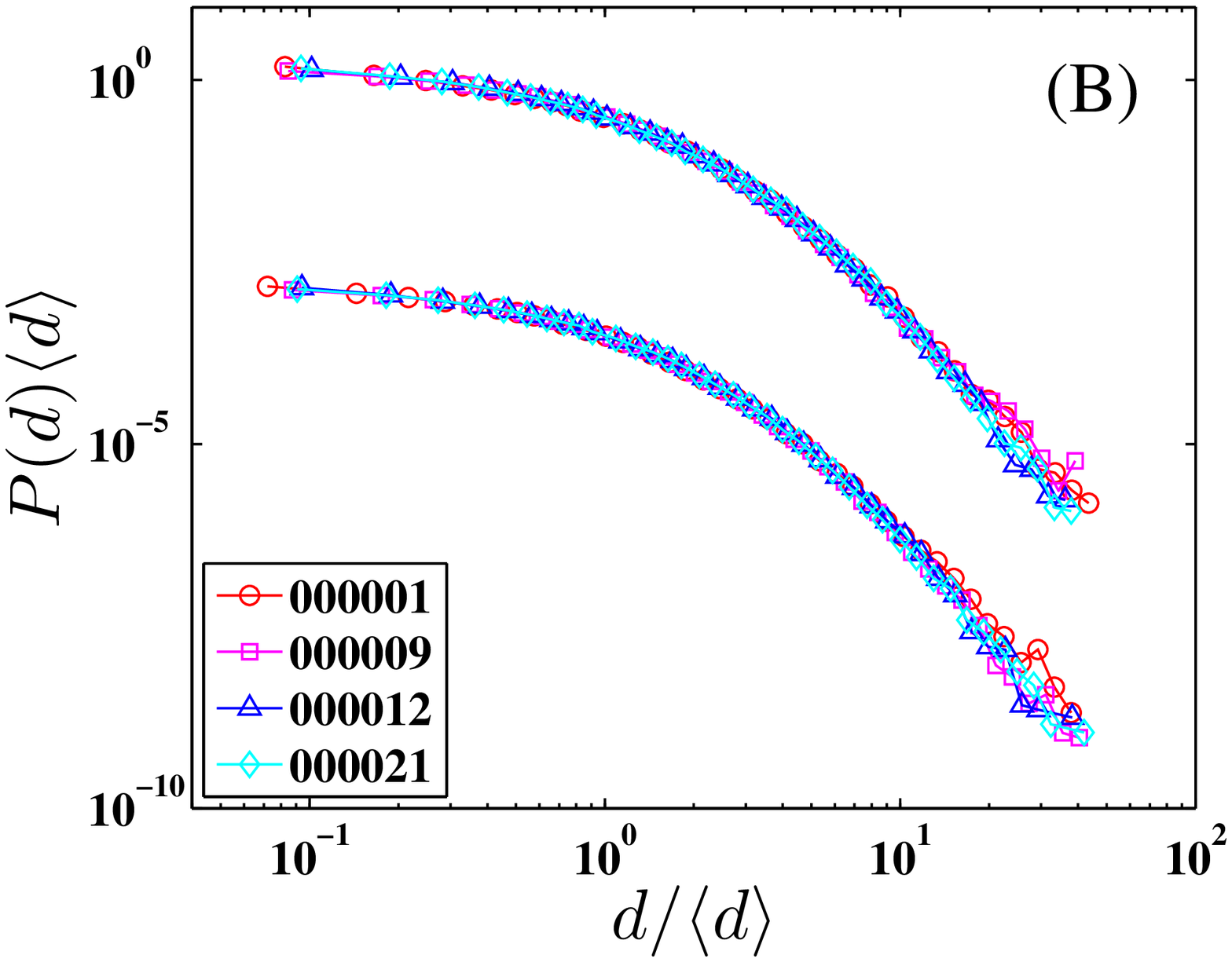}
  \caption{\label{Fig:PDF-4Stocks} Probability distributions (a) and rescaled probability distributions (b) of inter-cancellation durations $d$ for both cancelled buy and sell orders of four stocks 000001, 000009, 000012 and 000021. The curves corresponding to cancelled sell orders have been vertically translated downward for clarity.}
\end{figure}

Probability distribution of financial variables has crucial implications on asset pricing and risk management. In this section, we focus on investigating the probability distributions of inter-cancellation durations for both cancelled buy and sell cancelled orders of 18 stocks. The probability density functions (PDFs) $P(d)$ of four randomly chosen stocks are presented in Figure~\ref{Fig:PDF-4Stocks}(a).

According to the empirical results shown in the literature and the curve shape presented in Figure \ref{Fig:PDF-4Stocks}, we fit the distributions by Weibulls \cite{Eisler-Kertesz-2006-EPJB,Sazuka-2007-PA,Jiang-Chen-Zhou-2008-PA,Ni-Jiang-Gu-Ren-Chen-Zhou-2010-PA}
and $q$-exponentials \cite{Jiang-Chen-Zhou-2008-PA,Poloti-Scalas-2008-PA,Ni-Jiang-Gu-Ren-Chen-Zhou-2010-PA,Nadarajah-Kotz-2006-PLA,Nadarajah-Kotz-2007-PA}. For Weibull distributions, we have
\begin{equation}
P_{\rm{WBL}}(d)=\frac{b}{a}\left(\frac{d}{a}\right)^{b-1}{\rm{e}}^{-(\frac{d}{a})^b},
\label{Eq:WBL}
\end{equation}
where $a$ is the scale parameter and $b$ is the shape parameter. The $q$-exponential distributions can be described as follows:
\begin{equation}
P_{\rm{qE}}(d)=\frac{1}{\kappa}\left[1-(1-q)(\frac{d}{\kappa})\right]^{\frac{q}{1-q}},
\label{Eq:qE}
\end{equation}
where $\kappa$ is the scale parameter and $q$ is the shape parameter.

The maximum likelihood estimation (MLE) method is first applied to estimate the parameters of Weibull and $q$-exponential distributions. As we know, MLE method captures the major part of the fitting data. We find that it accounts for $64.9\%$ ($63.7\%$) in the range $d \leq 10$ for cancelled buy (sell) orders. The parameters $a$ and $b$ of the Weibull distribution and the parameters $\kappa$ and $q$ of the $q$-exponential distribution estimated with the MLE method are listed in the left panel of Table \ref{Tb:PDF_W_q_b}.

Since Weibull distribution has the same number of parameters as $q$-exponential distribution, the root mean square $\chi$ of the difference between the best fit and the empirical data is applied to compare the performance of the two distributions, which is presented in Table \ref{Tb:PDF_W_q_b} as well. It is clear that $\chi_{\rm{qE}}$ of $q$-exponential distribution are smaller than $\chi_{\rm{WBL}}$ of Weibull distribution ($\chi_{\rm{qE}} < \chi_{\rm{WBL}}$). So we conclude that $q$-exponential distribution outperforms Weibull distribution for the cancelled buy orders of each stock with the MLE method.

\setlength\tabcolsep{3.5pt}
\begin{table}[htp]
  \centering
  \caption{Parameters of the Weibull and $q$-exponential distributions based on the MLE and NLSE methods for cancelled buy orders of 18 stocks. $\chi$ is the root mean square of the difference between the best fit and the empirical data.}
  \label{Tb:PDF_W_q_b}
  \centering
  \begin{tabular}{ccrcccrcccrcccrcc}
  \hline \hline
  && \multicolumn{7}{c}{MLE} && \multicolumn{7}{c}{NLSE} \\
  \cline{3-9} \cline{11-17}
  Stock && \multicolumn{3}{c}{Weibull} && \multicolumn{3}{c}{$q$-exponential} && \multicolumn{3}{c}{Weibull} && \multicolumn{3}{c}{$q$-exponential} \\
  \cline{3-5} \cline{7-9} \cline{11-13} \cline{15-17}
  && $a$~~~ & $b$ & $\chi_{\rm{WBL}}$ && $\kappa~~~$ & $q$ & $\chi_{\rm{qE}}$ && $a~~~$ & $b$ & $\chi_{\rm{WBL}}$ && $\kappa~~~$ & $q$ & $\chi_{\rm{qE}}$ \\
  \hline
    000001 && 11.55 & 0.92 & 0.0075 && 9.47 & 1.22 & 0.0057 && 4.88 & 0.54 & 0.0092 && 7.97 & 1.26 & 0.0029 \\
    000009 && 11.45 & 0.94 & 0.0064 && 9.59 & 1.18 & 0.0044 && 4.73 & 0.54 & 0.0114 && 7.34 & 1.28 & 0.0013 \\
    000012 && 9.57 & 0.95 & 0.0090 && 8.00 & 1.18 & 0.0066 && 4.66 & 0.59 & 0.0098 && 7.07 & 1.21 & 0.0041 \\
    000016 && 11.63 & 0.96 & 0.0073 && 9.99 & 1.16 & 0.0055 && 7.26 & 0.68 & 0.0062 && 9.11 & 1.17 & 0.0038 \\
    000021 && 10.31 & 0.93 & 0.0085 && 8.52 & 1.20 & 0.0063 && 4.96 & 0.58 & 0.0092 && 7.67 & 1.22 & 0.0042 \\
    000024 && 11.56 & 0.93 & 0.0082 && 9.64 & 1.20 & 0.0065 && 6.44 & 0.63 & 0.0068 && 8.46 & 1.22 & 0.0043 \\
    000066 && 10.32 & 0.95 & 0.0082 && 8.67 & 1.18 & 0.0060 && 5.52 & 0.62 & 0.0088 && 7.62 & 1.21 & 0.0034 \\
    000406 && 11.59 & 0.95 & 0.0069 && 9.85 & 1.17 & 0.0052 && 4.35 & 0.54 & 0.0108 && 7.43 & 1.26 & 0.0016 \\
    000429 && 12.38 & 0.96 & 0.0067 && 10.74 & 1.15 & 0.0051 && 7.66 & 0.68 & 0.0064 && 9.29 & 1.18 & 0.0026 \\
    000488 && 12.24 & 0.88 & 0.0099 && 9.67 & 1.27 & 0.0087 && 6.83 & 0.60 & 0.0051 && 9.25 & 1.24 & 0.0079 \\
    000539 && 12.16 & 0.76 & 0.0216 && 7.91 & 1.51 & 0.0211 && 6.31 & 0.52 & 0.0161 && 8.75 & 1.36 & 0.0222 \\
    000541 && 12.69 & 0.92 & 0.0086 && 10.60 & 1.20 & 0.0075 && 6.01 & 0.59 & 0.0070 && 8.60 & 1.24 & 0.0049 \\
    000550 && 10.38 & 0.90 & 0.0102 && 8.32 & 1.24 & 0.0082 && 6.19 & 0.62 & 0.0067 && 8.14 & 1.22 & 0.0078 \\
    000581 && 12.63 & 0.84 & 0.0113 && 9.60 & 1.32 & 0.0106 && 6.05 & 0.54 & 0.0065 && 8.67 & 1.31 & 0.0094 \\
    000625 && 10.63 & 0.89 & 0.0105 && 8.37 & 1.27 & 0.0086 && 6.17 & 0.61 & 0.0066 && 8.47 & 1.22 & 0.0087 \\
    000709 && 12.78 & 0.94 & 0.0061 && 10.88 & 1.17 & 0.0046 && 5.37 & 0.56 & 0.0104 && 8.34 & 1.26 & 0.0016 \\
    000720 && 11.28 & 0.63 & 0.0394 && 4.57 & 2.00 & 0.0346 && 7.82 & 0.51 & 0.0369 && 9.24 & 1.47 & 0.0429 \\
    000778 && 13.69 & 0.94 & 0.0066 && 11.71 & 1.17 & 0.0056 && 7.77 & 0.65 & 0.0061 && 10.25 & 1.19 & 0.0038 \\
    MEAN   && 12.01 & 0.89 & 0.0104 && 9.56 & 1.27 & 0.0088 && 6.50 & 0.60 & 0.0092 && 8.79 & 1.25 & 0.0076 \\
  \hline \hline
 \end{tabular}
\end{table}

In order to capture the tail behavior of the distribution, we then utilize the nonlinear least-squares estimation (NLSE) method to fit the distribution of cancelled buy orders, and the parameters of Weibull and $q$-exponential distributions are listed in the right panel of Table \ref{Tb:PDF_W_q_b}. The parameters $a$ and $b$ calculated from the NLSE method are all smaller than those with the MLE method for 18 stocks. For the parameter $\kappa$, 3 stocks out of 18 stocks own larger values with NLSE method, while for the parameter $q$, 12 stocks have larger values with NLSE method. We also select the root mean square $\chi$ to compare the performance of the two distributions with the NLSE method. According to the values of $\chi$ listed in Table \ref{Tb:PDF_W_q_b}, we find that the result is different from the MLE method. There are 6 stocks prefer Weibull distribution, and the rest 12 stocks are better fitted by $q$-exponential distribution. The mean values of four parameters for cancelled buy orders are also presented in the last row of Table \ref{Tb:PDF_W_q_b}. The mean value of the four parameters obtained from the MLE method are larger than those from the NLSE method.

With the same procedure mentioned above, we then analyze the probability distribution of cancelled sell orders with the MLE and NLSE methods, and obtain similar results. The parameters $a$ and $b$ of Weibull distribution and the parameters $\kappa$ and $q$ of $q$-exponential distribution are listed in Table \ref{Tb:PDF_W_q_s}. For the cancelled sell orders, the relation $\chi_{\rm{qE}} < \chi_{\rm{WBL}}$ is satisfied for each stock when using the MLE method, which indicates that the distribution prefers $q$-exponential distribution than Weibull distribution with the MLE method. However, there are 3 stocks out of 18 stocks have smaller values of $\chi_{\rm{WBL}}$ and prefer Weibull distribution with the NLSE method.

\begin{table}[htp]
  \centering
  \caption{Parameters of Weibull and $q$-exponential distributions based on the MLE and NLSE methods for cancelled sell orders of 18 stocks. $\chi$ is the root mean square of the difference between the best fit and the empirical data.}
  \label{Tb:PDF_W_q_s}
  \centering
  \begin{tabular}{ccrcccrcccrcccrcc}
  \hline \hline
  && \multicolumn{7}{c}{MLE} && \multicolumn{7}{c}{NLSE} \\
  \cline{3-9} \cline{11-17}
  Stock && \multicolumn{3}{c}{Weibull} && \multicolumn{3}{c}{$q$-exponential} && \multicolumn{3}{c}{Weibull} && \multicolumn{3}{c}{$q$-exponential} \\
  \cline{3-5} \cline{7-9} \cline{11-13} \cline{15-17}
  && $a$~~~ & $b$ & $\chi_{\rm{WBL}}$ && $\kappa~~~$ & $q$ & $\chi_{\rm{qE}}$ && $a~~~$ & $b$ & $\chi_{\rm{WBL}}$ && $\kappa~~~$ & $q$ & $\chi_{\rm{qE}}$ \\
  \hline
    000001 && 13.05 & 0.90 & 0.0058 && 10.68 & 1.22 & 0.0042 && 6.04 & 0.56 & 0.0094 && 9.01 & 1.27 & 0.0015 \\
    000009 && 11.11 & 0.95 & 0.0066 && 9.44 & 1.17 & 0.0046 && 4.92 & 0.57 & 0.0108 && 7.97 & 1.22 & 0.0015 \\
    000012 && 10.31 & 0.95 & 0.0083 && 8.64 & 1.18 & 0.0061 && 4.92 & 0.59 & 0.0094 && 7.58 & 1.21 & 0.0036 \\
    000016 && 11.96 & 0.97 & 0.0063 && 10.52 & 1.13 & 0.0047 && 5.12 & 0.59 & 0.0109 && 7.81 & 1.22 & 0.0021 \\
    000021 && 10.72 & 0.95 & 0.0073 && 9.09 & 1.17 & 0.0053 && 4.63 & 0.57 & 0.0107 && 7.49 & 1.23 & 0.0016 \\
    000024 && 11.16 & 0.99 & 0.0065 && 9.83 & 1.12 & 0.0046 && 4.32 & 0.56 & 0.0122 && 6.80 & 1.25 & 0.0035 \\
    000066 && 10.65 & 0.95 & 0.0072 && 9.11 & 1.16 & 0.0052 && 5.80 & 0.63 & 0.0090 && 7.95 & 1.19 & 0.0025 \\
    000406 && 11.88 & 0.95 & 0.0078 && 10.23 & 1.16 & 0.0065 && 5.04 & 0.57 & 0.0088 && 7.88 & 1.24 & 0.0036 \\
    000429 && 12.87 & 0.98 & 0.0060 && 11.45 & 1.12 & 0.0044 && 9.44 & 0.77 & 0.0045 && 10.56 & 1.12 & 0.0029 \\
    000488 && 11.96 & 0.89 & 0.0125 && 9.72 & 1.24 & 0.0119 && 8.67 & 0.70 & 0.0064 && 10.13 & 1.17 & 0.0124 \\
    000539 && 13.63 & 0.86 & 0.0093 && 10.68 & 1.28 & 0.0088 && 7.30 & 0.57 & 0.0053 && 9.24 & 1.31 & 0.0073 \\
    000541 && 12.86 & 0.94 & 0.0075 && 11.11 & 1.16 & 0.0065 && 4.69 & 0.52 & 0.0102 && 7.41 & 1.31 & 0.0047 \\
    000550 && 10.06 & 0.94 & 0.0078 && 8.48 & 1.17 & 0.0057 && 4.22 & 0.55 & 0.0111 && 6.66 & 1.25 & 0.0016 \\
    000581 && 12.43 & 0.91 & 0.0080 && 10.32 & 1.21 & 0.0070 && 5.70 & 0.56 & 0.0075 && 7.88 & 1.29 & 0.0041 \\
    000625 && 10.33 & 0.94 & 0.0086 && 8.70 & 1.18 & 0.0067 && 3.71 & 0.53 & 0.0108 && 6.70 & 1.25 & 0.0024 \\
    000709 && 13.24 & 0.95 & 0.0056 && 11.48 & 1.15 & 0.0043 && 5.22 & 0.55 & 0.0106 && 8.51 & 1.25 & 0.0023 \\
    000720 && 14.37 & 0.66 & 0.0275 && 7.02 & 1.82 & 0.0259 && 9.87 & 0.53 & 0.0241 && 10.93 & 1.46 & 0.0304 \\
    000778 && 12.90 & 0.96 & 0.0068 && 11.36 & 1.13 & 0.0055 && 8.27 & 0.71 & 0.0050 && 10.02 & 1.15 & 0.0036 \\
    MEAN   && 11.98 & 0.93 & 0.0085 && 9.94 & 1.20 & 0.0070 && 6.02 & 0.59 & 0.0095 && 8.29 & 1.25 & 0.0048 \\
  \hline \hline
 \end{tabular}
\end{table}

Similar to the cancelled buy orders, the parameters $a$ and $b$ of Weibull distribution calculated from the NLSE method are smaller than those from the MLE method for cancelled sell orders of each stock. However, when considering $q$-exponential distribution, there are 2 stocks (000488 and 000720) having larger values of $\kappa$ and smaller values of $q$ with the NLSE method. The mean values of four parameters for the cancelled sell orders are presented in the last row of Table \ref{Tb:PDF_W_q_s} as well. It is evident that the mean value of the four parameters obtained from the MLE method are larger than those from the NLSE method, except for the parameter $q$.

\begin{figure}[htb]
\centering
\includegraphics[width=7cm]{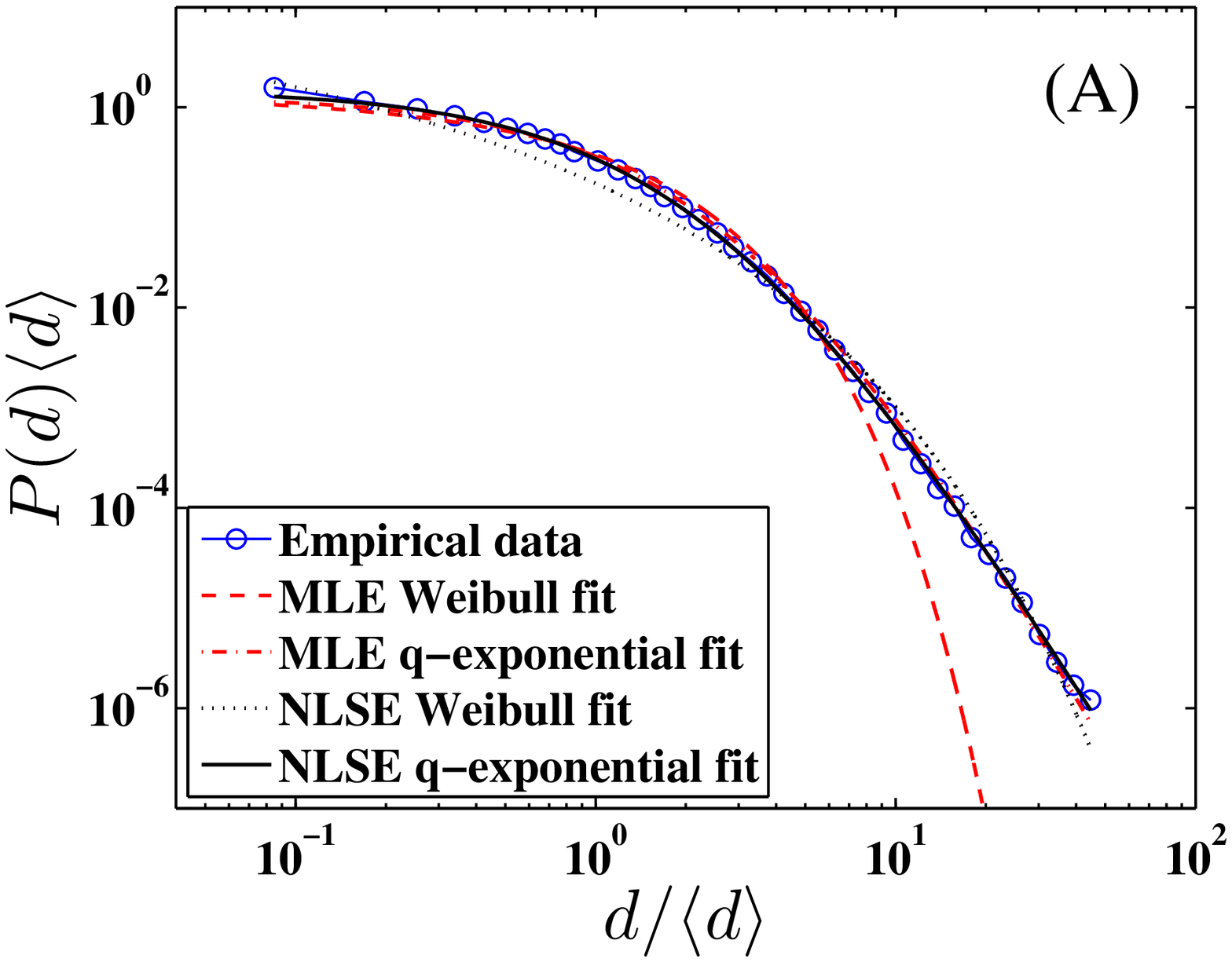}
\includegraphics[width=7cm]{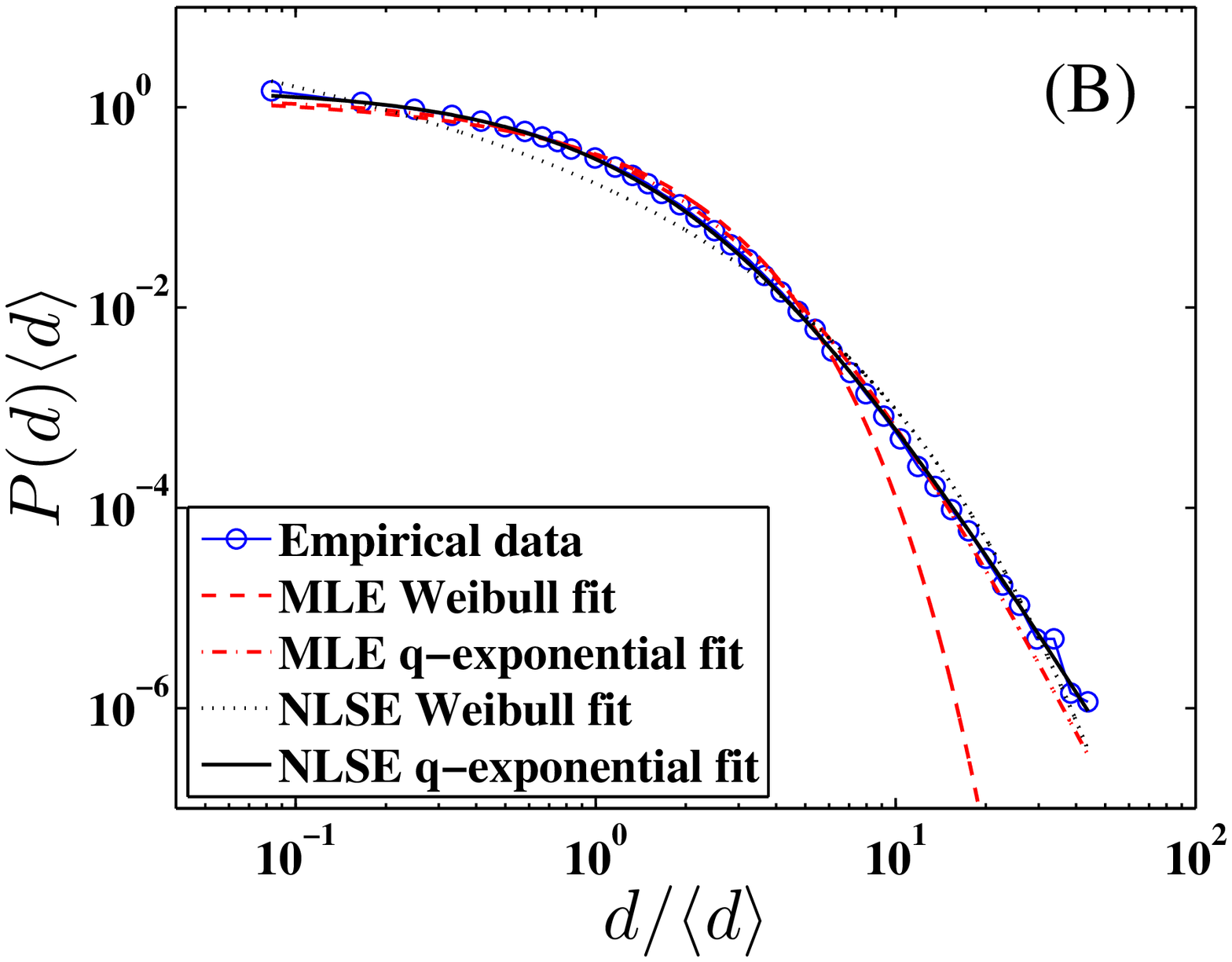}
\caption{\label{Fig:PDF-All} Rescaled probability distributions of the ensemble inter-cancellation durations for both cancelled buy (a) and sell (b) orders. The empirical data are fitted by Weibull and $q$-exponential distributions using the MLE and NLSE methods, respectively.}
\end{figure}

We rescale the inter-cancellation duration $d$ to $d/\langle{d}\rangle$ and the probability density function $P(d)$ to $P(d)\langle{d}\rangle$, where $\langle{d}\rangle$ is the mean value of inter-cancellation durations $d$. The rescaled PDFs of inter-cancellation durations for the same four stocks are presented in Figure \ref{Fig:PDF-4Stocks}(b). We find that four rescaled curves collapse together, showing a perfect scaling behavior. Since the rescaled probability distribution has an excellent scaling, we aggregate all the inter-cancellation durations of 18 stocks together and treat them as an ensemble to obtain a better statistic. The rescaled PDFs of ensemble durations for both cancelled buy and sell orders are shown in Figure \ref{Fig:PDF-All}.

The parameters of Weibull and $q$-exponential distributions for ensemble inter-cancellation durations are calculated by the MLE and NLSE methods, respectively. When Using the MLE method, we obtain $a=11.21,~b=0.91$ and $\kappa=9.13,~q=1.22$ for cancelled buy orders, and $a=11.55,~b=0.93$ and $\kappa=9.64,~q=1.19$ for cancelled sell orders. When applying the NLSE method, we have $a=4.79,~b=0.54$ and $\kappa=7.90,~q=1.25$ for cancelled buy orders, and $a=4.80,~b=0.54$ and $\kappa=7.91,~q=1.25$ for cancelled sell orders. It is clear that the values of the parameters obtained from ensemble inter-cancellation durations are similar to the mean values presented in the last rows of Table \ref{Tb:PDF_W_q_b} and Table \ref{Tb:PDF_W_q_s}, which confirms that the scaling behavior is truly existed. In addition, we find in Figure \ref{Fig:PDF-All} that Weibull distribution evidently deviates from the empirical data in the tail with the MLE method, which is consistent with the relation $\chi_{\rm{qE}} < \chi_{\rm{WBL}}$ for both cancelled buy and sell orders.

\section{Memory effect}
\label{se:memory}

Another important issue about financial time series is the memory effect. Many methods have been proposed for quantitatively measuring the memory effort, such as the rescaled range (RS) analysis \cite{Hurst-1951-TASCE,Mandelbrot-Ness-1968-SIAMR}, the fluctuation analysis (FA) \cite{Peng-Buldyrev-Goldberger-Havlin-Sciortino-Simons-Stanley-1992-Nature}, the wavelet transform module maxima (WTMM) method \cite{Holschneider-1988-JSP,Muzy-Bacry-Arneodo-1991-PRL}, the detrended fluctuation analysis (DFA) \cite{Peng-Buldyrev-Havlin-Simons-Stanley-Goldberger-1994-PRE}, the detrending moving average (DMA) \cite{Alessio-Carbone-Castelli-Frappietro-2002-EPJB}, and so on. Shao et al. compared the performance of the FA, DFA, and DMA methods using different long-range correlated time series, and found that centred detrending moving average (CDMA) has the best performance and DFA is only slightly worse in some situations, while FA performs the worst \cite{Shao-Gu-Jiang-Zhou-Sornette-2012-SR}. In this paper we apply the DFA and CDMA to investigate the memory effect of inter-cancellation duration series for both cancelled buy and sell orders of 18 stocks.

\begin{figure}[htb]
\centering
\includegraphics[width=7cm]{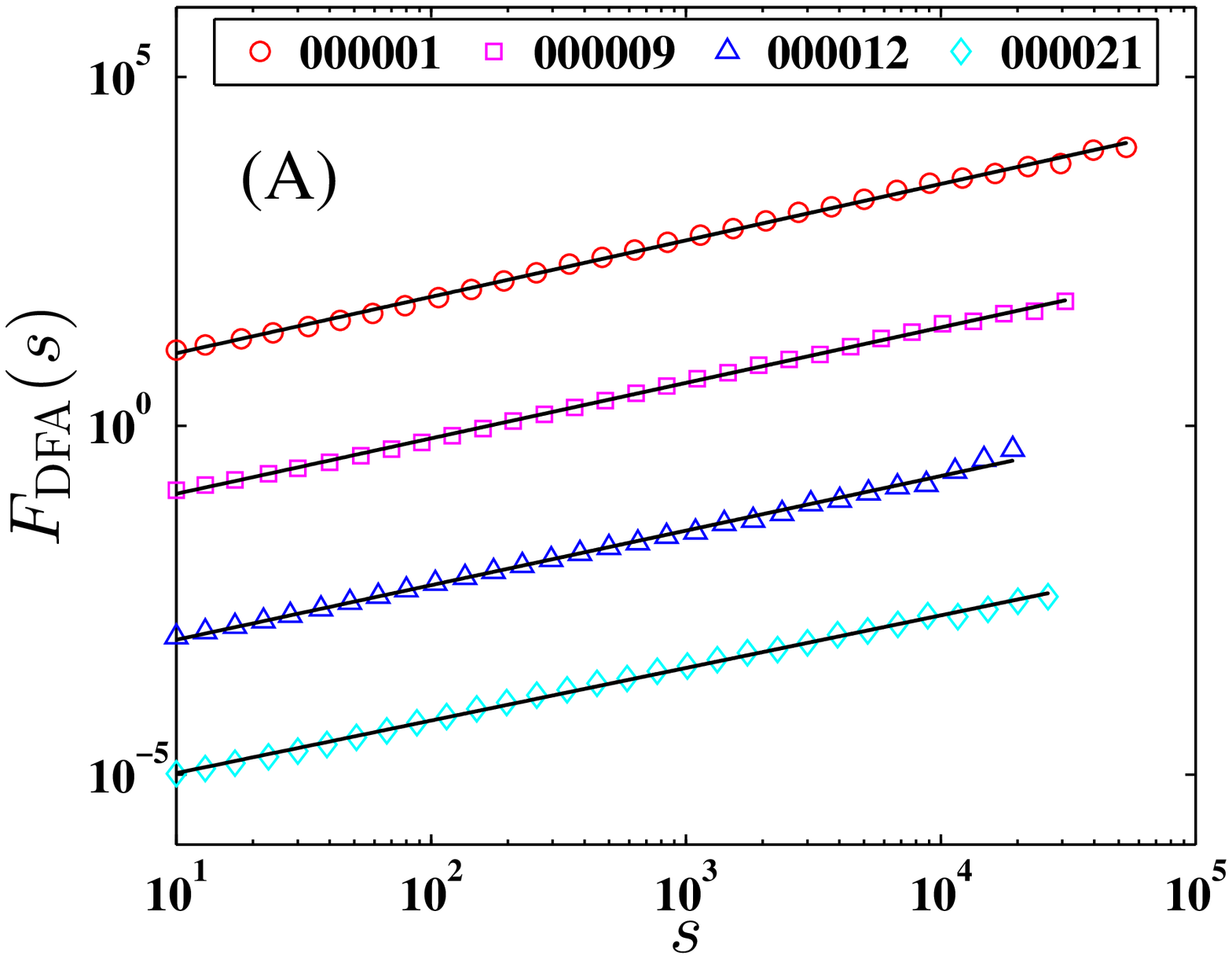}
\includegraphics[width=7cm]{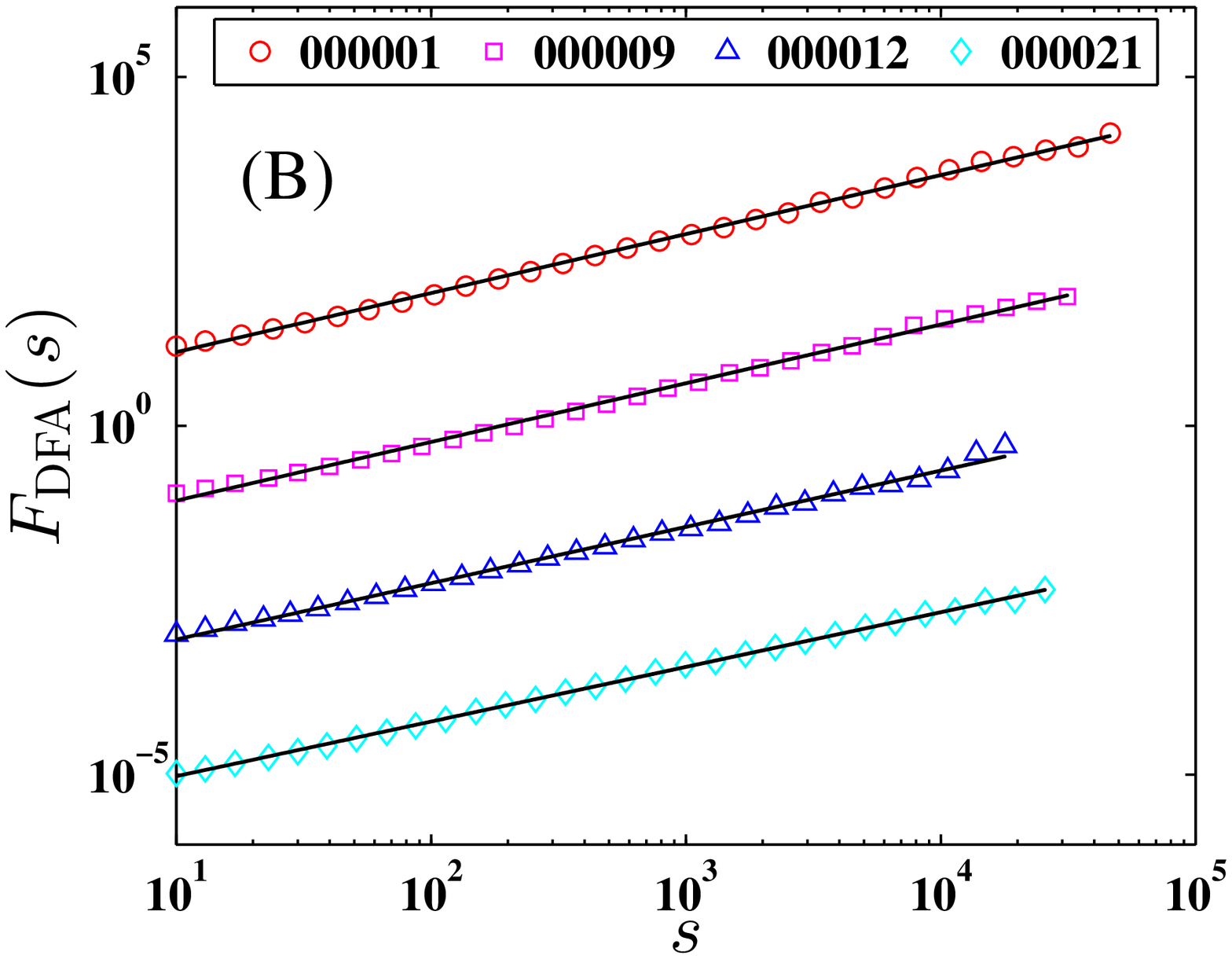}
\caption{\label{Fig:DFA-4Stocks} Plots of the fluctuation functions $F_{\rm{DFA}}(s)$ of inter-cancellation durations for both cancelled buy (a) and sell (b) orders for four stocks with the DFA method. The solid lines are power-law fits to the empirical data. The curves of stocks 000009, 000012 and 000021 have been shifted vertically for clarity.}
\end{figure}

Figure \ref{Fig:DFA-4Stocks} presents the detrended fluctuation functions $F_{\rm{DFA}}(s)$ with respect to the size scale $s$ using the DFA method for both cancelled buy and sell orders of four stocks, 000001, 000009, 000012 and 000021. Each curve reveals excellent power-law scaling over more than three orders of magnitude.

Applying the DFA method, the Hurst exponents $H$ of 18 stocks are estimated according to the power-law relation $F_{\rm{DFA}}(s)\sim{s}^H$, which are the slopes of solid lines shown in the log-log plot of Figure \ref{Fig:DFA-4Stocks}. We list the Hurst exponents for both cancelled buy and sell orders of 18 stocks in Table \ref{Tb:DFA-DMA-H}. The value of $H$ for cancelled buy orders varies in the range $[0.68, 0.82]$ with the mean value $\langle{H}\rangle=0.76 \pm 0.04$, and for cancelled sell orders it varies from 0.68 to 0.85 with the mean value $\langle{H}\rangle=0.76 \pm 0.04$. Since all the Hurst exponents are evidently larger than 0.5, we conclude that the inter-cancellation duration time series of both cancelled buy and sell orders process long memory.

\begin{table}[htp]
  \centering
  \caption{Hurst exponents $H$ of inter-cancellation durations for both cancelled buy and sell orders of 18 stocks based on the DFA and DMA methods. $H_{\rm{SFL}}$ is the mean Hurst exponent of 100 shuffled inter-cancellation durations.}
  \medskip
  \label{Tb:DFA-DMA-H}
  \centering
  \begin{tabular}{ccccccccccccc}
  \hline \hline
  && \multicolumn{5}{c}{Cancelled buy orders} && \multicolumn{5}{c}{Cancelled sell orders} \\
  \cline{3-7} \cline{9-13}
  Stock && \multicolumn{2}{c}{DFA} && \multicolumn{2}{c}{DMA} && \multicolumn{2}{c}{DFA} && \multicolumn{2}{c}{DMA} \\
  \cline{3-4} \cline{6-7} \cline{9-10} \cline{12-13}
  && $H$ & $H_{\rm{SFL}}$ && $H$ & $H_{\rm{SFL}}$ && $H$ & $H_{\rm{SFL}}$ && $H$ & $H_{\rm{SFL}}$ \\
  \hline
    000001 && 0.808 & 0.500 && 0.814 & 0.499 && 0.845 & 0.501 && 0.837 & 0.497 \\
    000009 && 0.795 & 0.501 && 0.796 & 0.495 && 0.841 & 0.502 && 0.839 & 0.494 \\
    000012 && 0.782 & 0.503 && 0.763 & 0.495 && 0.806 & 0.500 && 0.791 & 0.493 \\
    000016 && 0.739 & 0.500 && 0.726 & 0.495 && 0.726 & 0.500 && 0.737 & 0.499 \\
    000021 && 0.754 & 0.501 && 0.753 & 0.496 && 0.783 & 0.500 && 0.777 & 0.495 \\
    000024 && 0.773 & 0.501 && 0.768 & 0.492 && 0.731 & 0.503 && 0.746 & 0.497 \\
    000066 && 0.775 & 0.500 && 0.770 & 0.498 && 0.797 & 0.500 && 0.796 & 0.494 \\
    000406 && 0.734 & 0.500 && 0.738 & 0.496 && 0.783 & 0.500 && 0.784 & 0.494 \\
    000429 && 0.719 & 0.503 && 0.708 & 0.494 && 0.679 & 0.502 && 0.690 & 0.493 \\
    000488 && 0.764 & 0.501 && 0.751 & 0.492 && 0.747 & 0.501 && 0.721 & 0.498 \\
    000539 && 0.811 & 0.501 && 0.818 & 0.494 && 0.752 & 0.500 && 0.740 & 0.496 \\
    000541 && 0.741 & 0.502 && 0.722 & 0.503 && 0.715 & 0.502 && 0.712 & 0.496 \\
    000550 && 0.773 & 0.500 && 0.766 & 0.493 && 0.762 & 0.500 && 0.761 & 0.495 \\
    000581 && 0.816 & 0.500 && 0.809 & 0.496 && 0.768 & 0.501 && 0.757 & 0.500 \\
    000625 && 0.743 & 0.500 && 0.739 & 0.490 && 0.745 & 0.501 && 0.742 & 0.496 \\
    000709 && 0.747 & 0.500 && 0.742 & 0.496 && 0.738 & 0.501 && 0.731 & 0.500 \\
    000720 && 0.679 & 0.500 && 0.685 & 0.492 && 0.710 & 0.501 && 0.701 & 0.493 \\
    000778 && 0.721 & 0.502 && 0.719 & 0.494 && 0.724 & 0.503 && 0.717 & 0.494 \\
  \hline \hline
 \end{tabular}
\end{table}

\begin{figure}[htb]
\centering
\includegraphics[width=7cm]{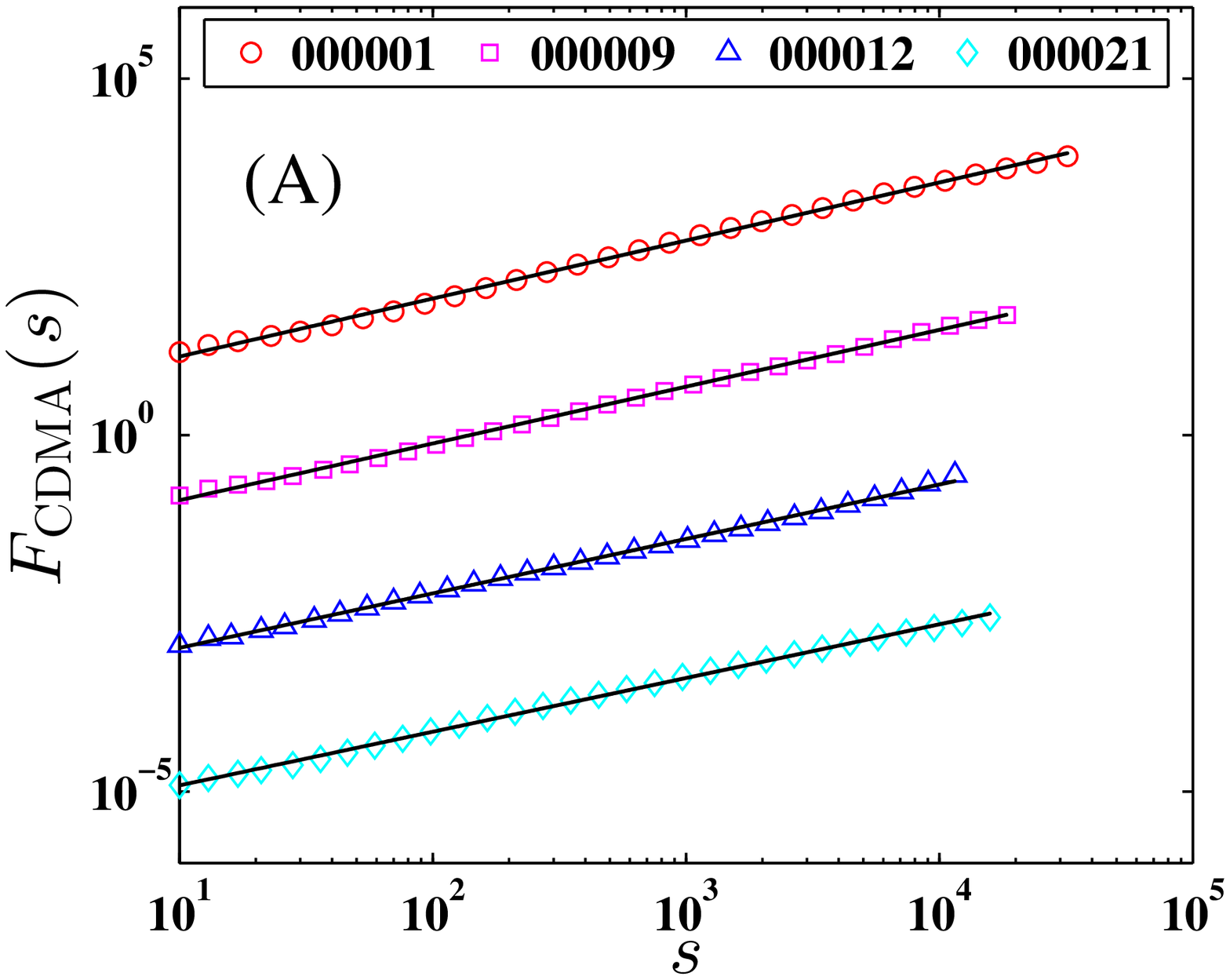}
\includegraphics[width=7cm]{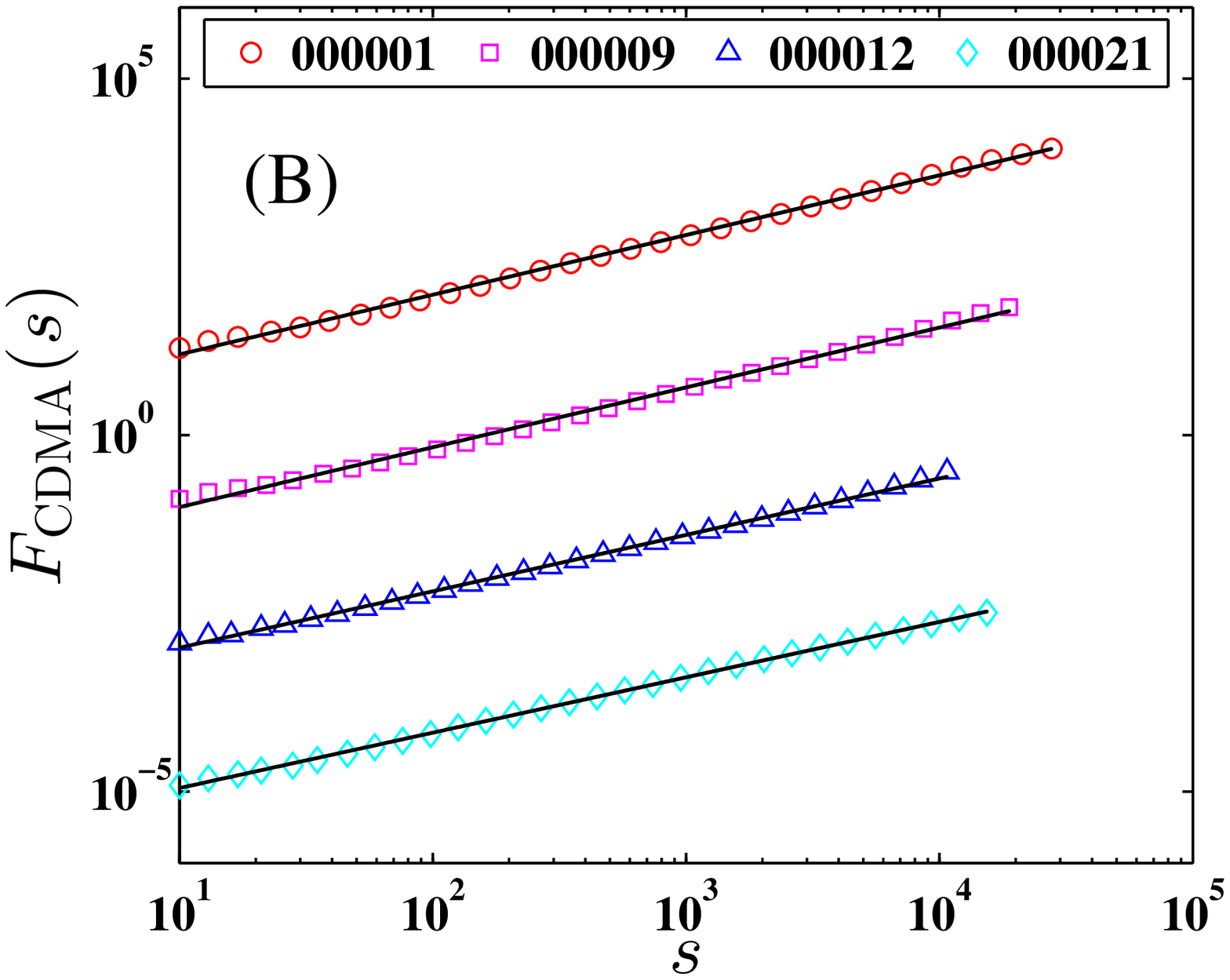}
\caption{\label{Fig:DMA-4Stocks} Plots of the fluctuation functions $F_{\rm{CDMA}}(s)$ of inter-cancellation durations for both cancelled buy (a) and sell (b) orders for four stocks with the CDMA method. The solid lines are power-law fits to the empirical data. The curves of stocks 000009, 000012 and 000021 have been shifted vertically for clarity.}
\end{figure}

Comparing with the backward detrending moving average (BDMA) and forward detrending moving average (FDMA), the centered detrending moving average (CDMA) performs better in one dimensional time series \cite{Arianos-Carbone-2007-PA}. We then choose the CDMA method to estimate the memory effect of inter-cancellation duration series. The detrended fluctuation functions $F_{\rm{CDMA}}(s)$ calculated from the CDMA method for both cancelled buy and sell orders of four stocks are depicted in Figure \ref{Fig:DMA-4Stocks}.

Perfect power-law scalings are observed in the log-log plot which implies that the relation $F_{\rm{CDMA}}(s)\sim{s}^H$ is well satisfied. The Hurst exponents $H$ are the slopes of solid lines in the log-log plot. Using the least-squares fitting method, we calculate the Hurst exponents for both cancelled buy and sell orders of 18 stocks which are listed in Table \ref{Tb:DFA-DMA-H} as well. It is obvious that the values of $H$ obtained from the CDMA method are close to those calculated from DFA method. With all the Hurst exponents apparently larger than 0.5 with the two methods, we conclude that the inter-cancellation duration time series for both cancelled buy and sell orders of 18 stocks process long memory.

On the other hand, the memory effect might be affected by the distribution of inter-cancellation durations. In order to test this hypothesis, we first shuffle the inter-cancellation duration series of each stock for 100 times, and then calculate the Hurst exponents $H_{\rm{SFL}}$ for each shuffling series based on both DFA and DMA methods, respectively. The mean Hurst exponents $H_{\rm{SFL}}$ of 100 shuffling series for both cancelled buy and sell orders of each stock are shown in Table \ref{Tb:DFA-DMA-H}. We find that the values of $H_{\rm{SFL}}$ are all extremely close to 0.5, being significant smaller than the original ones $H$. So we conclude that the distribution of inter-cancellation duration series has little impact on its memory effect, and confirm that inter-cancellation duration series of both cancelled buy and sell orders truly exhibit significant long memory for all the 18 stocks.

Memory effect presents the time persistence of inter-cancellation durations. It reflects the clustering behavior of order cancellation which is caused by traders' similar reactions to the market. For example, when good news arrives, traders will immediately cancel their limit orders in order to avoid being transacted at the unfavorable price. Many cancellations occur in a short period, which results to the clustering behavior and long memory effect of order cancellation.

\section{Multifractal nature}
\label{se:multifractal}

Multifractals are ubiquitous in the nature and society \cite{Mandelbrot-1983}, including financial time series. In this section, we investigate the multifractal properties of inter-cancellation durations applying the multifractal detrended fluctuation analysis (MF-DFA) method \cite{Kantelhardt-Zschiegner-KoscielnyBunde-Havlin-Bunde-Stanley-2002-PA} which is generalized from the DFA method. The MF-DFA algorithm is described as follows.

{\em{Step 1}}. Consider an inter-cancellation duration series $d(i)$, where $i=1,2,\cdots,N$. Construct the cumulative sum sequence $y(i)$ as follows,
\begin{equation}
 y(i)=\sum_{k=1}^{i}{d(k)}, ~~i=1, 2, \cdots, N.
 \label{Eq:DFA-y}
\end{equation}

{\em{Step 2}}. Divide the sequence $y(i)$ into $N_s$ disjoint segments with the same length $s$, where $N_s=[N/s]$, and $[x]$ is the largest integer not larger than $x$. Each segment can be denoted as $y_v$ such that $y_v(j)=y(\ell+j)$ for $1\leqslant{j}\leqslant{s}$, and $\ell=(v-1)s$. Since the length of the inter-cancellation duration series $N$ might not be a multiple of the segment size $s$, a remaining part (with the length smaller than $s$) at the end of sequence $y(i)$ is not covered by the dividing procedure. We will select another $N_s$ disjoint segments from the end of the series for compensating the remaining part, and then consider the $2N_s$ segments which covers the whole sequence $y(i)$.

{\em{Step 3}}. In each segment $y_v$, a polynomial function is utilized to represent the trend by the least-squares regression. The simplest function could be a line, and in the paper we adopt the linear function $\widetilde{y}_v(j)$ with $1\leqslant{j}\leqslant{s}$ to remove the trend. The residual $\epsilon_{v}(j)$ in the segment $y_v$ can be calculated by
\begin{equation}
\epsilon_{v}(j)=y_{v}(j)-\widetilde{y}_{v}(j)~.
 \label{Eq:DFA-epsilon}
\end{equation}

{\em{Step 4}}. The detrended fluctuation function $F(v,s)$ of the segment $y_v$ is defined as follows,
\begin{equation}
F(v,s) = \sqrt{\frac{1}{s}\sum_{j=1}^{s}[\epsilon_{v}(j)]^2}~.
 \label{Eq:DFA-F1}
\end{equation}

{\em{Step 5}}. The $q$-th order overall fluctuation function $F_q(s)$ is determined through
\begin{equation}
F_q(s)=\left\{\frac{1}{2N_s}\sum_{v=1}^{2N_s} {[F(v,s)]^q}\right\}^{\frac{1}{q}},
 \label{Eq:MFDFA-Fq}
\end{equation}
where $q$ can take any real value except for $q=0$. When $q=0$, according to L'H\^{o}spital's rule we have
\begin{equation}
\ln[F_0(s)] = \frac{1}{2N_s}\sum_{v=1}^{2N_s}{\ln[F(v,s)]}.
 \label{Eq:MFDFA-Fq0}
\end{equation}

{\em{Step 6}}. Varying the values of segment size $s$ from 10 to $[N/6]$, a power-law relation between the function $F_q(s)$ and the size scale $s$ is determined, which reads
\begin{equation}
F_q(s) \sim s^{h(q)}.
 \label{Eq:MFDFA-h}
\end{equation}

According to the standard multifractal formalism, the multifractal scaling exponent $\tau(q)$ characterizes the multifractal nature, which reads
\begin{equation}
\tau(q)=qh(q)-D_f,
 \label{Eq:MFDFA-tau}
\end{equation}
where $D_f$ is the fractal dimension of the geometric support of multifractal measure. For one dimensional time series analysis, we have $D_f=1$. If the scaling exponent $\tau(q)$ is a nonlinear function of $q$, the series has multifractal nature. Finally, it is easy to obtain the singularity strength function $\alpha(q)$ and the multifractal spectrum $f(\alpha)$ via the Legendre transform, that is,
\begin{equation}
\left\{
 \begin{array}{ll}
  \alpha(q)={\rm{d}}\tau(q)/{\rm{d}}q\\
  f(q)=q{\alpha}-{\tau}(q)
 \end{array}
\right..
 \label{Eq:MFDFA-alphaf}
\end{equation}

\begin{figure}[htb]
\centering
\includegraphics[width=7cm]{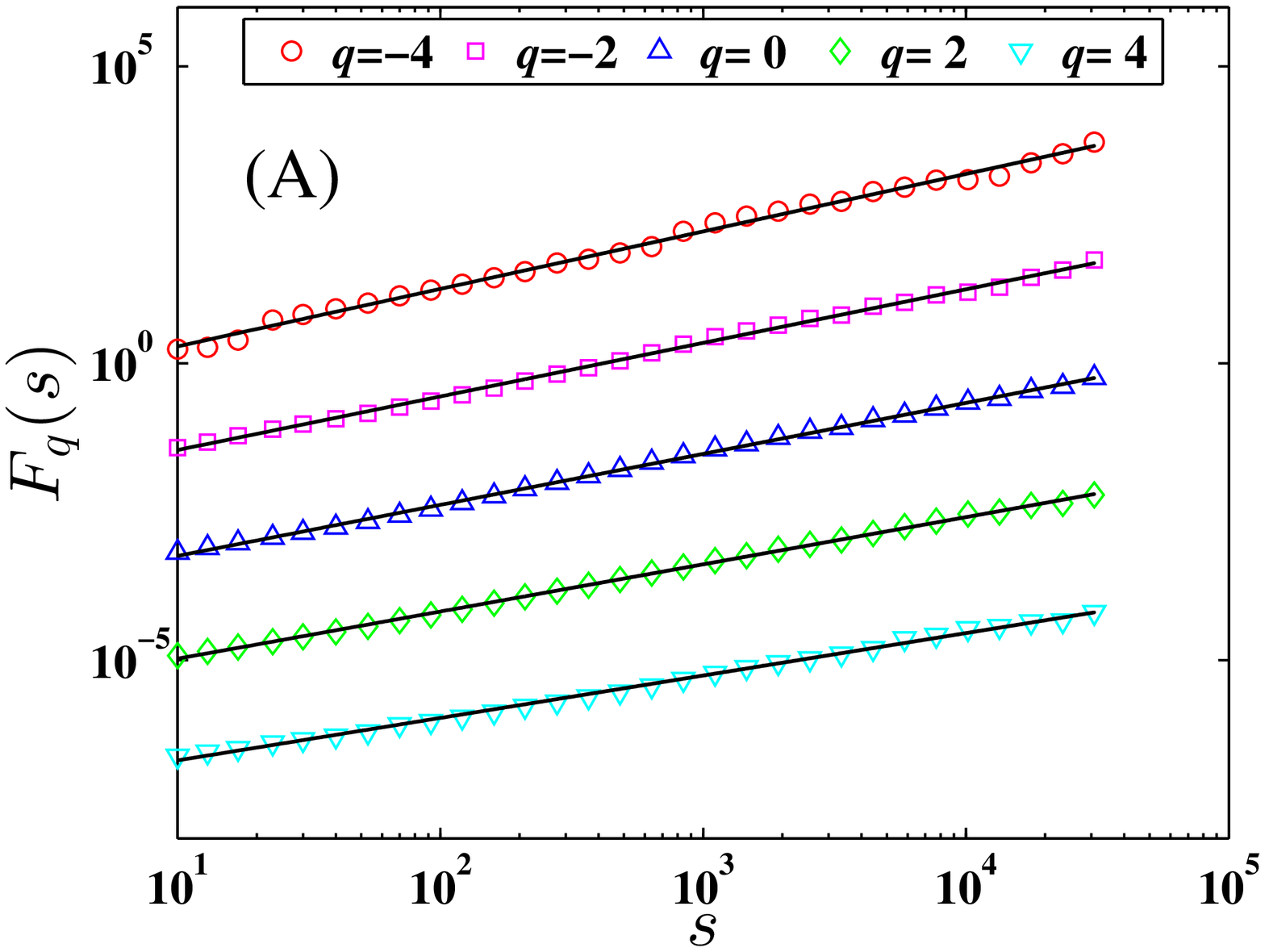}
\includegraphics[width=7cm]{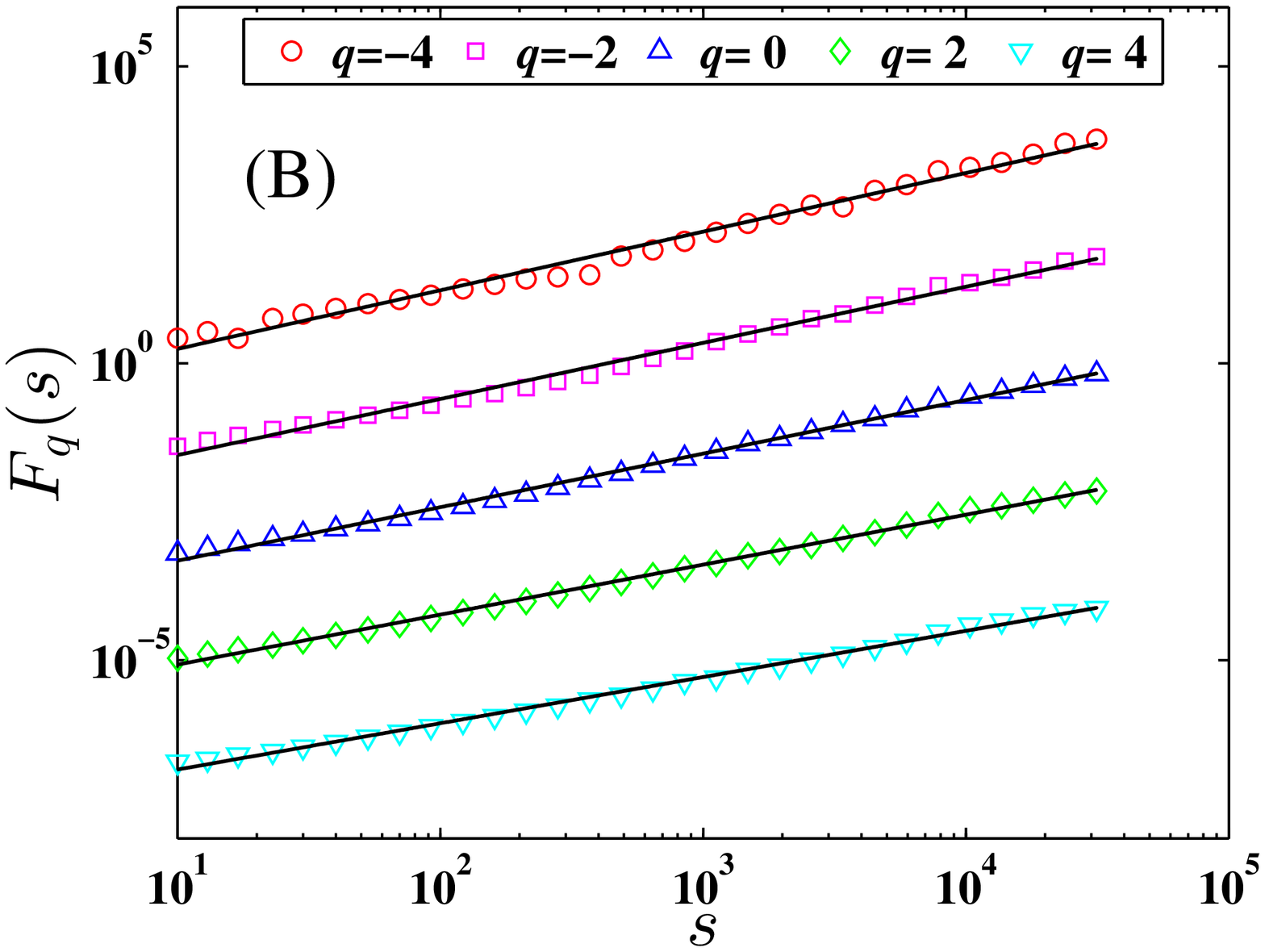}
\includegraphics[width=7cm]{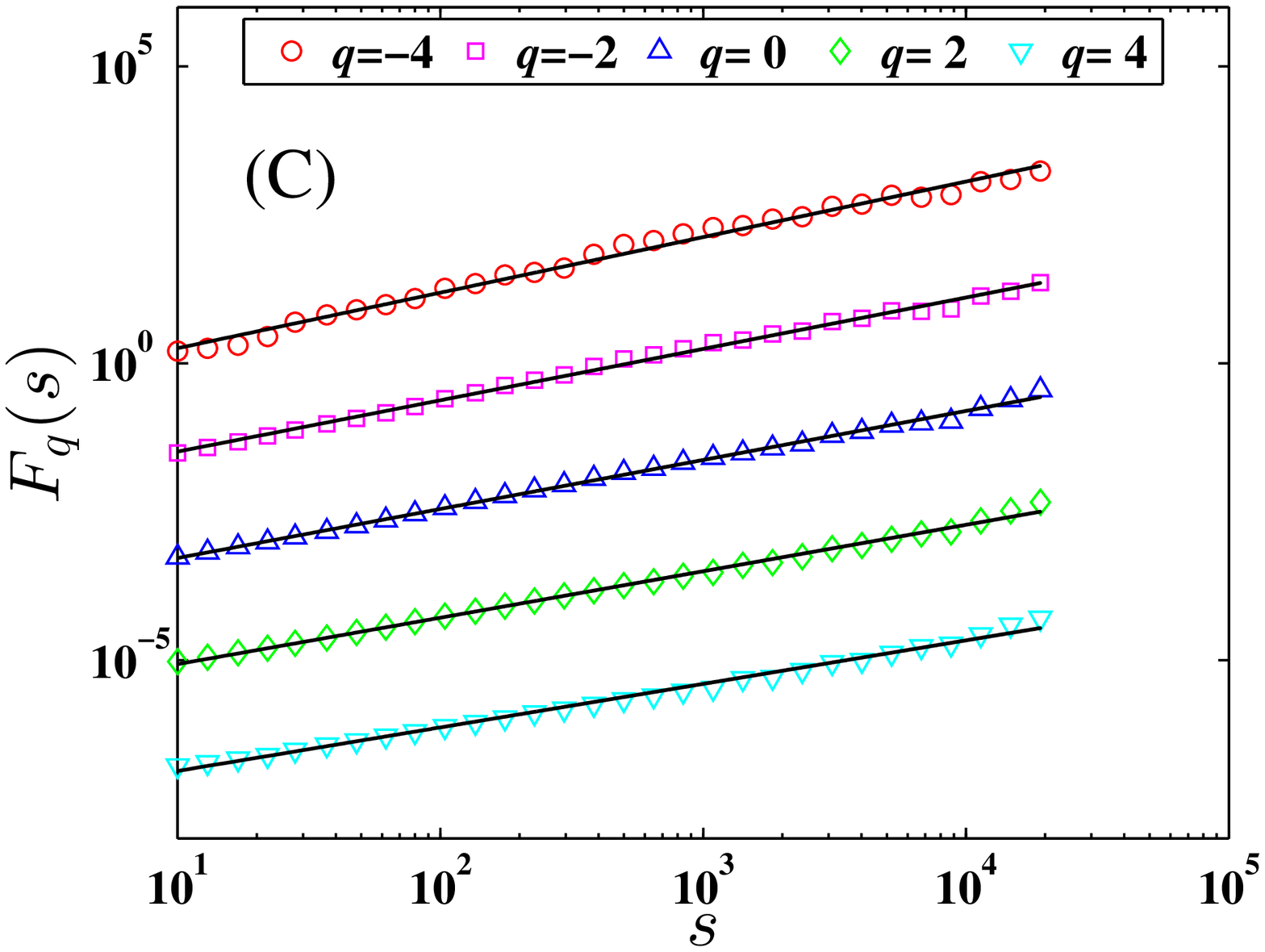}
\includegraphics[width=7cm]{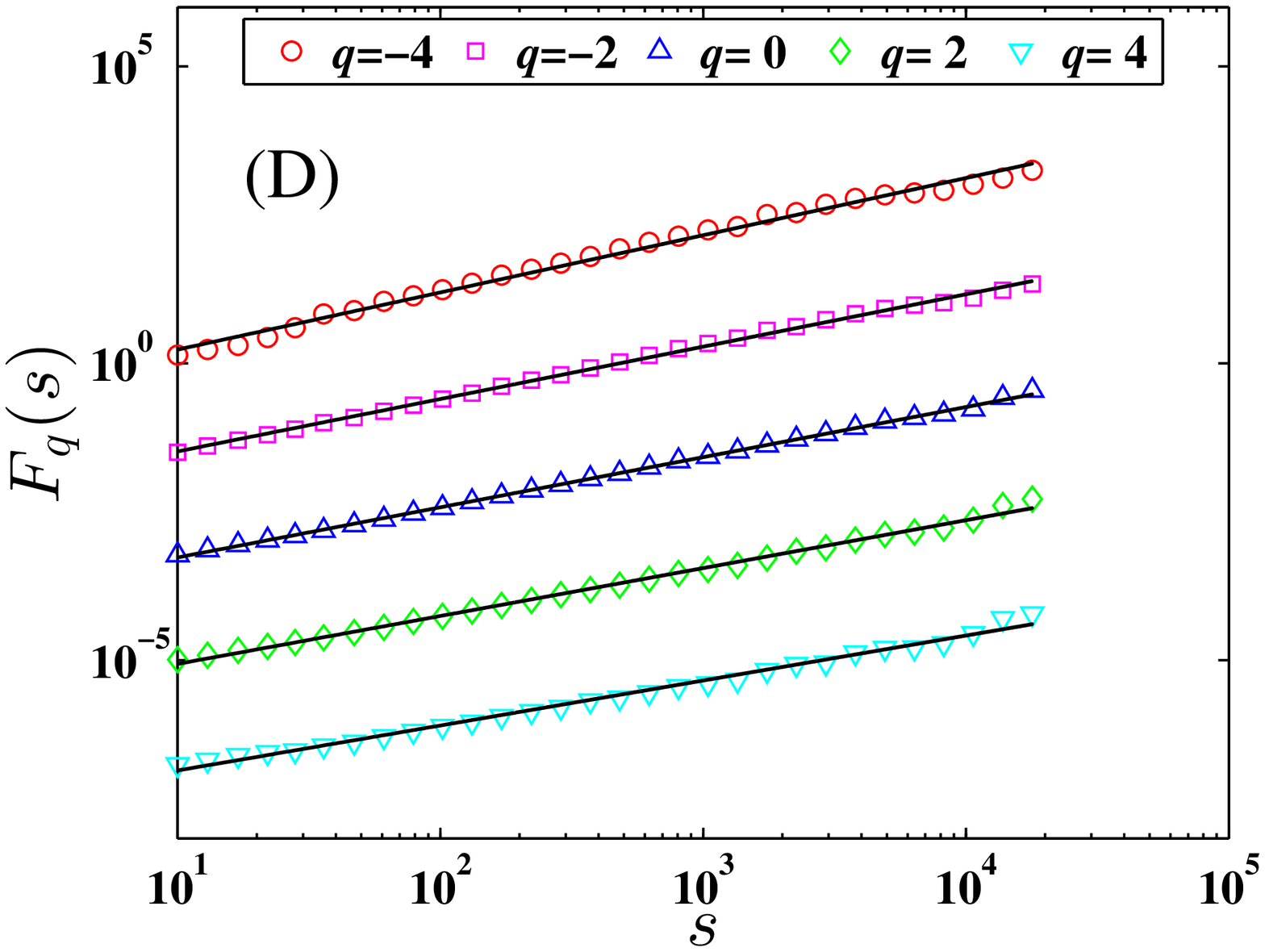}
\caption{\label{Fig:MFDFA-Fq} Plots of the $q$-th order detrended fluctuation functions $F_q(s)$ for the cancelled buy (a) and sell (b) orders of the stock 000009, and for the cancelled buy (c) and sell (d) orders of the stock 000012. The solid lines are the best power-law fits to the data. The plots for $q=-2,0,2,4$ have been translated downward for better visibility.}
\end{figure}

We calculate the $q$-th order fluctuation functions $F_q(s)$ of inter-cancellation durations for both cancelled buy and sell orders of two stocks, 000009 and 000012, and present the fluctuation functions $F_q(s)$ in Figure \ref{Fig:MFDFA-Fq}. We find that the function $F_q(s)$ has a excellent power-law scaling with respect to the scale size $s$. Using the least-squares fitting method, we obtain the slopes $h(q)$ for $q=-4,-2,0,2,4$, respectively.

\begin{figure}[htb]
\centering
\includegraphics[width=7cm]{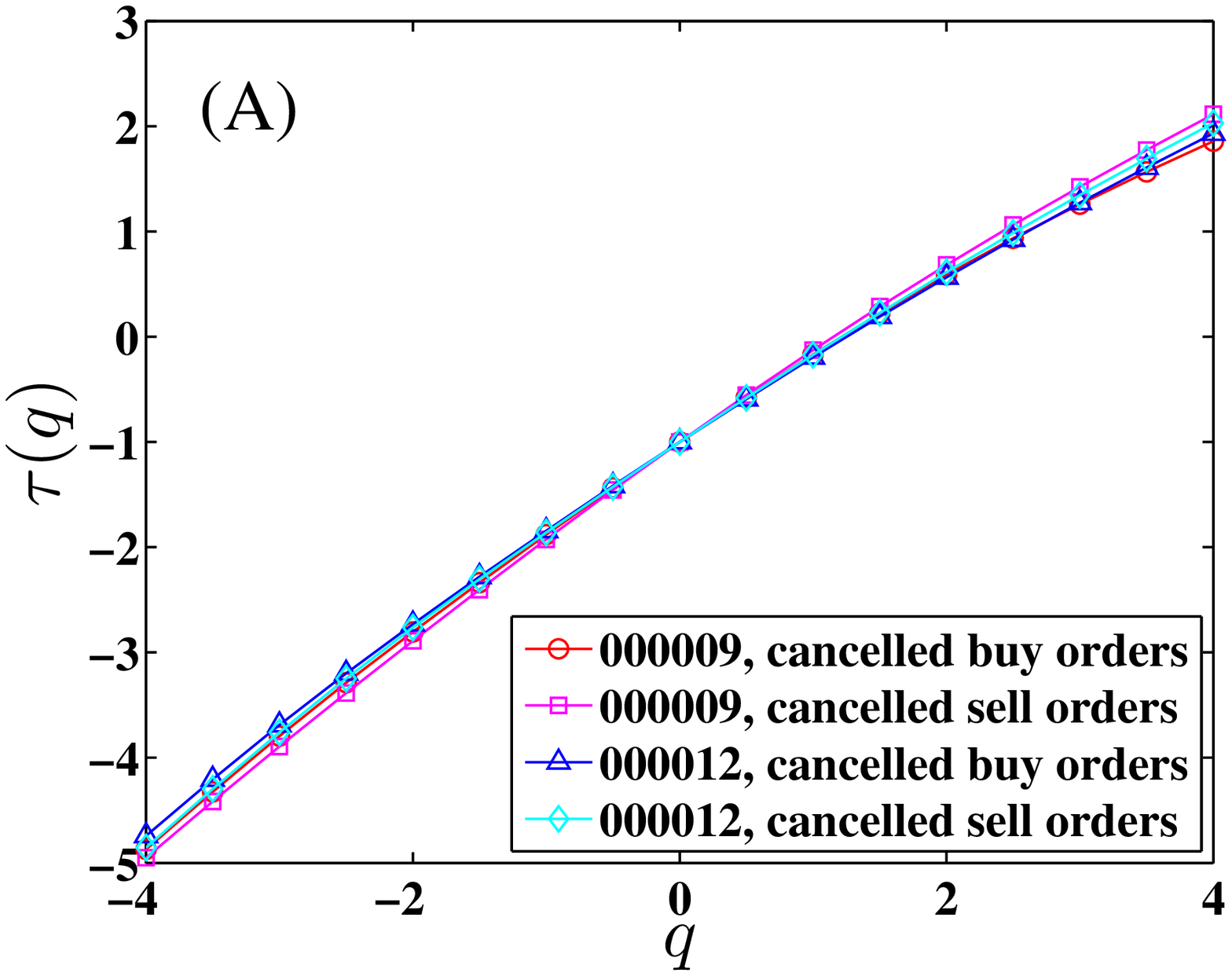}
\includegraphics[width=7cm]{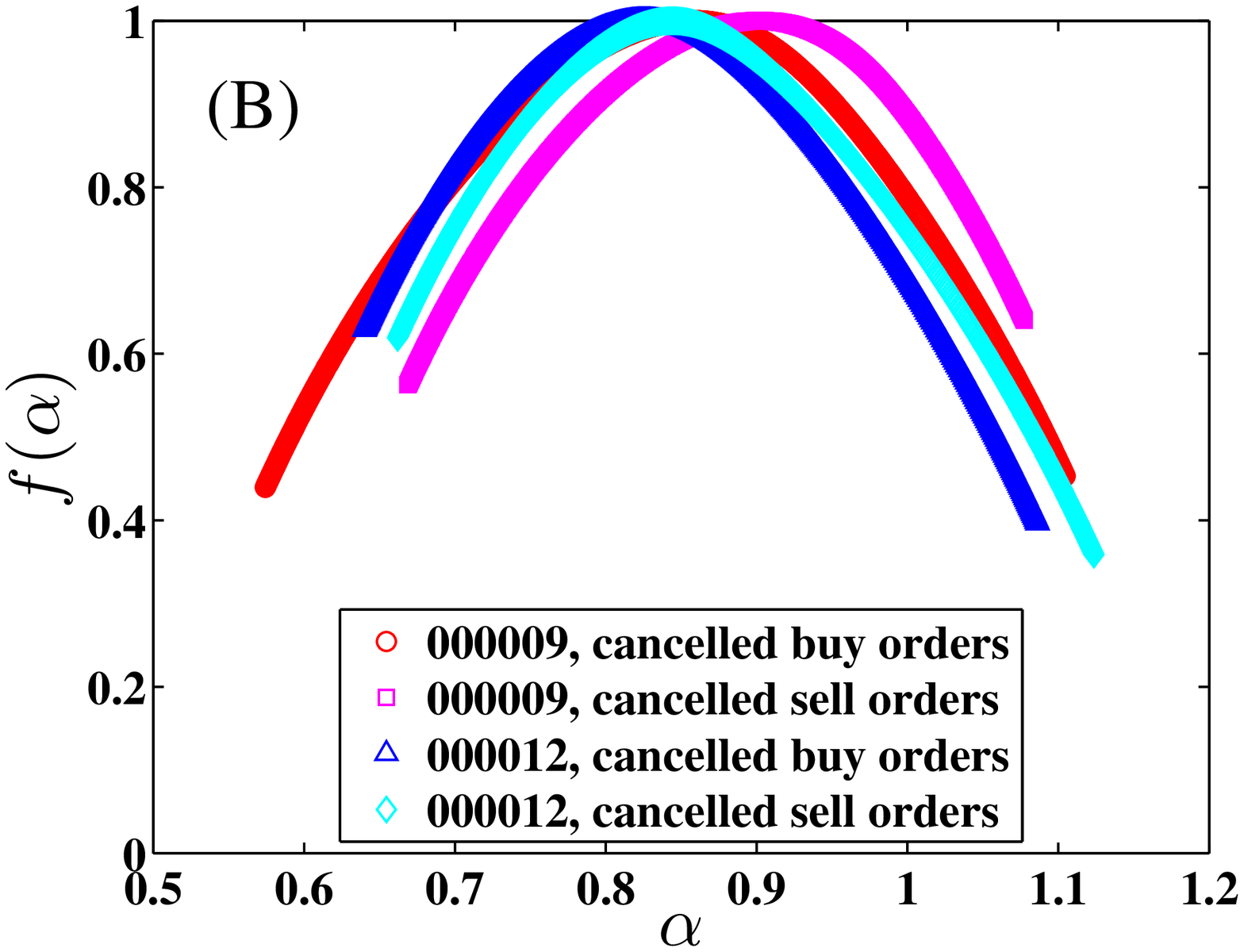}
\caption{\label{Fig:MFDFA-tau-alpha-f} Plots of the scaling exponents $\tau(q)$ (a) and multifractal spectra $f(\alpha)$ (b) for both cancelled buy and sell orders of the stocks 000009 and 000012.}
\end{figure}

Figure \ref{Fig:MFDFA-tau-alpha-f} presents the scaling exponents $\tau(q)$ with respect to the order $q$ and the multifractal spectra $f(\alpha)$ as a function of the singularity strength $\alpha$ for both cancelled buy and sell orders of the two stocks. We observe that the function $\tau(q)$ is nonlinear with respect to $q$, which illustrates that the inter-cancellation durations own multifractal nature.

In addition, the strength of multifractal can also be measured by the width of the multifractal spectrum $f(\alpha)$  ($\Delta\alpha=\alpha_{\rm{max}}-\alpha_{\rm{min}}$), and a larger value of $\Delta\alpha$ corresponds to stronger multifractal. We calculate the singularity width $\Delta\alpha$ for both cancelled buy and sell orders of 18 stocks and list the results in Table \ref{Tb:MFDFA-dalpha}. The value $\Delta\alpha$ of cancelled buy orders varies from 0.37 to 1.28 with the mean value $\langle{\Delta\alpha}\rangle=0.74 \pm 0.24$, and for cancelled sell orders the value $\Delta\alpha$ varies in the range $[0.41,1.38]$ with the mean value $\langle{\Delta\alpha}\rangle=0.68 \pm 0.25$. Since all the values of $\Delta\alpha$ larger than zero, we consider that the inter-cancellation duration series for both cancelled buy and sell orders of 18 stocks have multifractal nature, which is consistent with the results obtained from the scaling exponent $\tau(q)$.

\begin{table}[htp]
  \centering
  \caption{The width of the multifractal spectra $\Delta\alpha$ of inter-cancellation durations for both cancelled buy and sell orders of 18 stocks based on the MF-DFA method. $\Delta\alpha_{\rm{SFL}}$ is the mean width of 100 shuffled inter-cancellation durations. $R$ is the residual of spectrum width by removing the shuffled width $\Delta\alpha_{\rm{SFL}}$ from the original one $\Delta\alpha$.}
  \medskip
  \label{Tb:MFDFA-dalpha}
  \centering
  \begin{tabular}{ccccccccc}
  \hline \hline
  && \multicolumn{3}{c}{Cancelled buy orders} && \multicolumn{3}{c}{Cancelled sell orders} \\
  \cline{3-5} \cline{7-9}
  Stock && $\Delta\alpha$ & $\Delta\alpha_{\rm{SFL}}$ & $R$ && $\Delta\alpha$ & $\Delta\alpha_{\rm{SFL}}$ & $R$ \\
  \hline
    000001 && 0.81 & 0.29 $\pm$ 0.01 & 0.52 && 0.80 & 0.29 $\pm$ 0.01 & 0.51 \\
    000009 && 0.53 & 0.32 $\pm$ 0.01 & 0.21 && 0.41 & 0.27 $\pm$ 0.02 & 0.13 \\
    000012 && 0.45 & 0.30 $\pm$ 0.02 & 0.15 && 0.46 & 0.30 $\pm$ 0.02 & 0.16 \\
    000016 && 0.37 & 0.26 $\pm$ 0.02 & 0.10 && 0.78 & 0.32 $\pm$ 0.02 & 0.46 \\
    000021 && 0.67 & 0.28 $\pm$ 0.02 & 0.39 && 0.50 & 0.28 $\pm$ 0.02 & 0.22 \\
    000024 && 0.97 & 0.32 $\pm$ 0.03 & 0.65 && 0.56 & 0.39 $\pm$ 0.02 & 0.17 \\
    000066 && 0.36 & 0.28 $\pm$ 0.02 & 0.08 && 0.41 & 0.26 $\pm$ 0.02 & 0.15 \\
    000406 && 0.84 & 0.37 $\pm$ 0.02 & 0.47 && 0.78 & 0.30 $\pm$ 0.02 & 0.48 \\
    000429 && 0.62 & 0.29 $\pm$ 0.03 & 0.33 && 0.41 & 0.24 $\pm$ 0.03 & 0.17 \\
    000488 && 0.77 & 0.36 $\pm$ 0.03 & 0.41 && 0.66 & 0.29 $\pm$ 0.03 & 0.37 \\
    000539 && 1.07 & 0.53 $\pm$ 0.04 & 0.54 && 0.92 & 0.43 $\pm$ 0.03 & 0.49 \\
    000541 && 0.70 & 0.44 $\pm$ 0.04 & 0.26 && 0.82 & 0.53 $\pm$ 0.04 & 0.29 \\
    000550 && 0.77 & 0.28 $\pm$ 0.02 & 0.49 && 0.58 & 0.33 $\pm$ 0.02 & 0.24 \\
    000581 && 0.84 & 0.47 $\pm$ 0.03 & 0.37 && 0.69 & 0.42 $\pm$ 0.04 & 0.26 \\
    000625 && 0.80 & 0.30 $\pm$ 0.02 & 0.50 && 0.61 & 0.34 $\pm$ 0.02 & 0.27 \\
    000709 && 0.87 & 0.34 $\pm$ 0.03 & 0.54 && 0.96 & 0.36 $\pm$ 0.03 & 0.59 \\
    000720 && 1.28 & 0.70 $\pm$ 0.06 & 0.58 && 1.38 & 0.60 $\pm$ 0.06 & 0.78 \\
    000778 && 0.62 & 0.31 $\pm$ 0.02 & 0.31 && 0.50 & 0.24 $\pm$ 0.02 & 0.25 \\
  \hline \hline
 \end{tabular}
\end{table}

Similar to the memory effect, the probability distribution might have influence upon the multifractal nature of inter-cancellation durations. In order to test the influence of distribution, we shuffle the inter-cancellation durations for 100 times to test this influence. For each shuffling series, the width of the multifractal spectrum $\Delta\alpha_{\rm{SFL}}$ is obtained based on the MF-DFA method. The mean values of 100 shuffled series for 18 stocks are listed in Table \ref{Tb:MFDFA-dalpha}. The values of $\Delta\alpha_{\rm{SFL}}$ are clearly larger than zero, which indicates that the distribution of inter-cancellation durations reliably generates multifractal. We define the residual of spectrum width $R$ through removing the shuffled width $\Delta\alpha_{\rm{SFL}}$ from the original one $\Delta\alpha$, that is, $R=\Delta\alpha-\Delta\alpha_{\rm{SFL}}$, and list the values $R$ of 18 stocks in Table \ref{Tb:MFDFA-dalpha}. Since the values of $R$ are evidently larger than zero, we conclude that inter-cancellation durations process multifractal nature for both cancelled buy and sell orders of all the stocks.

\section{Conclusion}
\label{se:conclusion}

Order cancellation is an important issue in the dynamics of price formation in financial markets. We have carried out empirical investigations on the statistical properties of inter-cancellation durations (in units of events) using the order flow data of 18 liquid stocks traded on the Shenzhen Stock Exchange in the whole year of 2003.

We first study the probability distributions of inter-cancellation durations for both cancelled buy and sell orders, and find that the rescaled probability density functions have a scaling behavior. When fitting the probability distributions by Weibull and $q$-exponential distributions, we find that both cancelled buy and sell orders prefer $q$-exponential distribution with MLE method. However, applying the NLSE method, we find that cancelled buy orders of 6 stocks and cancelled sell orders of 3 stocks prefer Weibull distribution which is different from the result obtained from the MLE method.

We then investigate the memory effect of inter-cancellation durations based on the detrended fluctuation analysis (DFA) and centered detrending moving average (CDMA) methods. Using the DFA method we obtain the average Hurst exponent of 18 stocks $\langle{H}\rangle=0.76$ for both cancelled buy and sell orders, and with the CDMA method it is $\langle{H}\rangle=0.75$ for both cancelled buy and sell orders. According to the results from these two methods, it is evident that the inter-cancellation duration series processes the same strength of long memory for both cancelled buy and sell orders.

Finally, we investigate the multifractal properties applying the multifractal detrended fluctuation analysis (MF-DFA) method. We find that the average width of multifractal spectrum $\langle{\Delta\alpha}\rangle = 0.74$ for cancelled buy orders of 18 stocks and it is $\langle{\Delta\alpha}\rangle = 0.68$ for cancelled sell orders. So we conclude that the inter-cancellation duration series has multifractal nature, and inter-cancellation duration series of buy orders has little stronger multifractality than cancelled sell orders. Our findings indicate that the cancellation process has a bursty behavior and possesses long-range correlations. Such non-Poisson behaviors have been unveils in many other human dynamics \cite{Jiang-Xie-Li-Podobnik-Zhou-Stanley-2013-PNAS}.

\bigskip

{\textbf{Acknowledgments:}}

This work was partly supported by National Natural Science Foundation of China (Grants No. 71101052, 71131007 and 11075054), Shanghai Rising Star (Follow-up) Program (Grant No. 11QH1400800), and the Fundamental Research Funds for the Central Universities.

\bibliography{E:/Papers/Auxiliary/Bibliography}

\end{document}